\documentclass{birkjour}
\usepackage{orcidlink}
\usepackage{xcolor}
\usepackage{url}
\usepackage{graphicx}
\usepackage{amsmath}
\usepackage{amsfonts}
\usepackage{physics}
\usepackage{hyperref} 
\usepackage{nicefrac}
\usepackage{doi}
\usepackage{slashed}
\usepackage{booktabs}
\usepackage[english]{babel}
\usepackage[autostyle, english = american]{csquotes}
\MakeOuterQuote{"}

\usepackage[style=numeric,sorting=none]{biblatex} 

\RequirePackage[all]{xy}

\theoremstyle{definition}

\theoremstyle{remark}

\numberwithin{equation}{section}

\newcommand{\BibTeX}{B\kern-0.1emi\kern-0.017emb\kern-0.15em\TeX}
\newcommand{\XYpic}{$\mathrm{X\kern-0.3em\raisebox{-0.18em}{Y}}$-$\mathrm{pic}\,$}

\newcommand{\cl}{C \kern -0.1em \ell}  

\newcommand{\STA}{\mathbb{G}_{1,3}}
\newcommand{\eSTA}{\mathbb{G}_{1,3}^+}
\newcommand{\APS}{\mathbb{G}_3}

\newcommand{\Spin}{\mathrm{Spin}}
\newcommand{\gm}{\gamma}
\newcommand{\sm}{\sigma}
\newcommand{\al}{\alpha}
\newcommand{\be}{\beta}

\newcommand{\Le}{\mathrm{L}}

\newcommand{\Fo}{\mathrm{F}}
\newcommand{\Ba}{\mathrm{B}}
\newcommand{\jj}{\mathrm{J}}
\newcommand{\kk}{\mathrm{K}}
\newcommand{\mm}{\mathrm{M}}
\newcommand{\nn}{\mathrm{N}}

\newcommand{\la}{\lambda}

\newcommand{\sip}{\!:\!}
\newcommand{\sop}{\mathbin{\hbox{$\wedge$}\kern-0.457em\raisebox{0ex}{$\cdot$}}}

\newcommand{\cproj}[1]{\left<#1\right>_{0\oplus3}}

\newcommand{\dagg}{\dagger}
\newcommand{\LP}[2]{\hat{B}\widetilde{\lambda}_#1\lambda_#2}
\newcommand{\mLP}[2]{\hat{B}^\dagger\lambda_#1^\dagger\lambda_#2^-}

\newcommand{\upSAb}[3]{\left[#1\right]^{#2 #3}}
\newcommand{\downSAb}[3]{\left[#1\right]_{#3 #2}}
\usepackage[normalem]{ulem}

\newcommand{\ed}{\end{document}}
\begin{document}

%
%
%
%
%
%
%
%
%

\title[Squared Massive Constructive Amplitudes]
 {The Scattering Algebra of Physical Space: Squared Massive Constructive Amplitudes}
\author[Moab Croft]{Moab Croft \orcidlink{0000-0003-4082-0428}}
%
\address{%
Department of Physics, Illinois State University\\
Normal\\
IL, 61790\\
USA}
%

\author[Neil Christensen]{Neil Christensen}
%
\address{%
Department of Physics, Illinois State University\\
Normal\\
IL, 61790\\
USA}
\email{nchris3@ilstu.edu}

%

%
%
\date{\today}
\dedicatory{Last Revised:\\ December 14, 2025}

\begin{abstract}
The \textit{Algebra of Physical Space} (APS) is used to explore the \textit{Constructive Standard Model} (CSM) of particle physics. Namely, this paper connects the spinor formalism of the APS to massive amplitudes in the CSM. A novel equivalency between traditional CSM and APS-CSM formalisms is introduced, called the \textit{Scattering Algebra} (SA), with example calculations confirming the consistency of results between both frameworks. Through this all, two significant insights are revealed: The identification of traditional CSM \textit{spin spinors} with \textit{Lorentz rotors} in the APS, and the connection of the CSM to various formalisms through \textit{ray spinor structure}. The CSM's results are replicated in massive cases, showcasing the power of the index-free, matrix-free, coordinate-free, geometric approach and paving the way for future research into massless cases, amplitude-construction, and Wigner little group methods within the APS.
\end{abstract}
\label{Abstract}

\maketitle
 
\newpage

\section{Introduction}\label{S0}

The \textit{Standard Model} (SM) of particle physics describes the elementary particles and their interactions. Traditionally, the SM uses fields which transform under the Lorentz group. Since the Lorentz representations have more degrees of freedom than the physical states of the particles, it necessitates the implementation of unphysical degrees of freedom. In order to cancel the resulting effects of these additional degrees of freedom, "gauged" versions of global symmetries must be implemented as well. In order to test the SM, experimental physicists build particle accelerators that collide these particles at high energy and measure the resulting particles and their properties. Theoretical physicists, on the other hand, calculate the probability of these collisions producing each final state. These probabilities are encoded in so-called scattering amplitudes, which are mathematical quantities representing the transition from an initial configuration to some new configuration (e.g. $e^++e^-\rightarrow\mu^++\mu^-$). However, the calculation of amplitudes in the SM is quite laborious and creates extremely complicated mathematical expressions. This point is more extensively discussed in \cite{Christensen2024}. This traditional method of calculation involves Feynman-diagrams, and to refer to this, the term \textit{Feynman-diagram Standard Model} (FSM) is adopted. Recently, an approach spawned from advances in \textit{S-Matrix Theory} and \textit{Twistor Theory} has led to the \textit{Constructive} Standard Model (CSM) of particle physics \cite{Christensen2018,Britto2005,ArkaniHamed2021,Ema2024,2Ema2024,Ema2025,Ni2025}. This CSM removes unnecessary degrees of freedom from the outset, only \textit{constructing} amplitudes and diagrams with the particles' relevant transformation groups: \textit{Wigner little groups}. This new approach is presently in agreement with the FSM's results in all $3$-point and $4$-point amplitudes \cite{Christensen2020,2Christensen2024,3Christensen2024,4Christensen2024}. Moreover, CSM diagrams are not simply a rewriting of FSM diagrams. As discussed in \cite{2Christensen2024}, many diagrams differ significantly from their more traditional FSM counterparts despite the full amplitude returning identical results when a momentum choice is made. It is worth noting that the methods developed here apply to field theories beyond the SM. However, we prefer to begin with the SM since it is renormalizable, it is well-tested with a large collection of validated scattering amplitudes, and is deeply grounded in physical reality.

A recent study explored the Wigner little group for photons within the geometric \textit{Spacetime Algebra}, $\STA$, demonstrating the utility of Geometric Algebra for studying Wigner little groups \cite{Croft2025}. Since the CSM deals directly with Wigner little groups, it is therefore reasonable to presume a similar utility of Geometric Algebra for studying the CSM. Indeed, this paper will demonstrate exactly this. The CSM approach will be presented using the geometric \textit{Algebra of Physical Space} (APS), $\APS$, as it most directly translates to the traditional approaches in CSM. However, this paper will not directly deal with Wigner little groups. That is a subject deserving of its own paper. Rather, this paper will connect the spinor formalism of the APS to the massive amplitudes of the CSM. It will be seen that the central (scalar plus pseudoscalar) projection of the APS replaces the matrix-trace to simplify squared amplitudes, and that the geometric perspective gives new insights into the CSM. The power of the Geometric Algebra approach is manifest throughout the paper: The structures of matrices and indices can be wholly explored with geometric, representation-free methods.  

The layout of this paper will be as follows: The APS and its conventions will be presented. Thereafter a spinor formalism will be presented, clarifying the geometry, terminology, and relationships between relevant concepts within the subject. Then the \textit{Scattering Algebra} (SA) will be introduced as an express equivalency between the traditional CSM methods and those of the APS. Finally, example calculations using this SA will be given by squaring the $h f\overline{f}$ (Higgs and two massive fermions) amplitude, the $W q\overline{q}$ (W-boson and two massive fermions) amplitude, the T-channel photon contribution to the $f_1\overline{f}_1\overline{f}_2f_2$ (two pairs of massive fermions) scattering amplitude, and also evaluating the S-channel Higgs-photon crossterm for the $f_1\overline{f}_1\overline{f}_2f_2$ scattering amplitude. In the appendix a short discussion on \textit{chirality} will be given, as well as a table of correspondence for the traditional and APS formalisms. Also in the appendix will be direct translations of all massive $2$-body decay amplitudes, all contributions to $f_1\overline{f}_1\overline{f}_2f_2$, and all contributions to $f\overline{f}hh$. Every amplitude presented in this paper has been confirmed to square to the correct value as given by both the CSM and FSM, and the authors of this paper believe the given examples are a sufficiently representative set such that the squaring of amplitudes not presented in this paper could be done following these rules.

\section{Essential Conventions}\label{A0}

The \textit{Algebra of Physical Space} (APS) is the real geometric algebra $\APS$ generated by the unipotent ($+1$-squaring) orthonormal vector basis
\begin{equation}\label{EQEC: Basis Vectors}
    \sm_x,\quad\sm_y,\quad\sm_z,
\end{equation}
which satisfies the inner product
\begin{equation}\label{EQEC: Vec. Inn. Prod.}
    \sm_j\cdot\sm_k=\frac{1}{2}(\sm_j\sm_k+\sm_k\sm_j)=\delta_{jk},
\end{equation}
where $\delta_{jk}$ is the Kronecker delta \cite{Doran2003,Hestenes1966,Baylis2004}. Note that the basis vectors are written with the same notation as the \textit{Pauli matrices}. This choice brings attention to the fact that those matrices act as a representation basis in $M_2(\mathbb{C})$ (algebra of $2\times2$ complex-entry matrices) for the APS:
\begin{equation}\label{EQEC: Pauli Mat.}
    1\equiv\begin{bmatrix}
        1 & 0 \\
        0 & 1
    \end{bmatrix},\,\,\sm_x\equiv\begin{bmatrix}
        0 & 1 \\
        1 & 0
    \end{bmatrix},\,\,\sm_y\equiv\begin{bmatrix}
        0 & -i \\
        i & 0
    \end{bmatrix},\,\,\sm_z\equiv\begin{bmatrix}
        1 & 0 \\
        0 & -1
    \end{bmatrix}.
\end{equation}
Any spatial vector $\mathbf{a}\in\APS$ can be written as a real linear combination of the vector basis,
\begin{equation}\label{EQEC: Vec. A}
    \mathbf{a}=\sum_ja_j\sm_j = a_x\sm_x+a_y\sm_y+a_z\sm_z.
\end{equation}
The geometric product of any two vectors $\mathbf{a}$ and $\mathbf{b}$ is given by 
\begin{equation}\label{EQEC: ab}
    \mathbf{a}\mathbf{b}=\mathbf{a}\cdot\mathbf{b}+\mathbf{a}\wedge\mathbf{b},
\end{equation}
and the inner and outer products are respectively given by
\begin{equation}\label{EQEC: a dot b}
    \mathbf{a}\cdot\mathbf{b}=\frac{1}{2}(\mathbf{a}\mathbf{b}+\mathbf{b}\mathbf{a}) = |\mathbf{a}||\mathbf{b}|\cos{\theta_{ab}}=\sum_ja_jb_j,
\end{equation}
and
\begin{equation}\label{EQEC: a wedge b}
    \mathbf{a}\wedge\mathbf{b}=\frac{1}{2}(\mathbf{a}\mathbf{b}-\mathbf{b}\mathbf{a})=|\mathbf{a}||\mathbf{b}|\sin{\theta_{ab}}\,\hat{a}\wedge\hat{b}=\frac{1}{2}\sum_{j\neq k}(a_jb_k-a_kb_j)\sm_{jk},
\end{equation}
where $\theta_{ab}$ is the angle between the two vectors in the plane that joins them, $|\mathbf{a}|$ and $|\mathbf{b}|$ are the magnitudes of the vectors as defined by EQ.~\ref{EQEC: Magnitude}, $\hat{a}=\mathbf{a}/|\mathbf{a}|$ and $\hat{b}=\mathbf{b}/|\mathbf{b}|$, and $\sm_{jk}=\sm_j\sm_k=\sm_j\wedge\sm_k$ are the unit bivectors for $j\neq k$. These unit bivectors form an anti-unipotent orthonormal basis,
\begin{equation}\label{EQEC: Biv. Basis}
    \sm_{xy}=\sm_x\sm_y,\quad\sm_{yz}=\sm_y\sm_z,\quad\sm_{zx}=\sm_z\sm_x.
\end{equation}
The APS also has a unit trivector, called the space's \textit{pseudoscalar},
\begin{equation}\label{EQEC: PSS}
    \sm_{xyz}=\sm_x\sm_y\sm_z=\sm_x\wedge\sm_y\wedge\sm_z.
\end{equation}
It can be easily shown that this pseudoscalar commutes with all elements of the APS, and is anti-unipotent. Therefore it is rewritten as $i=\sm_{xyz}$ and identified with the \textit{unit imaginary}. Through this, it is possible to interpret, and rewrite, bivectors as imaginary vectors,
\begin{equation}\label{EQEC: Imag. Vec.}
    i\sm_x=\sm_{yz},\quad i\sm_y=\sm_{zx},\quad i\sm_z=\sm_{xy}.
\end{equation}
Moreover, the unit scalar $1$ and pseudoscalar $i$ form the basis for the algebra's center, which is the subset whose elements commute with all elements of the APS, $\mathrm{Center}[\APS]\approx\mathbb{C}$, and is isomorphic to the algebra of complex numbers. 

A general element $g\in\APS$ is called a \textit{multivector}. It is a real linear combination of all bases, including $1$ and $i$, and the bases in EQ.~\ref{EQEC: Basis Vectors} and EQ.~\ref{EQEC: Biv. Basis}. If $\jj$ is the \textit{multivector index} for this complete algebraic basis, then
\begin{equation}\label{EQEC: Multivector}
   \begin{aligned}
        g=\sum_\jj g_\jj\sm_\jj &= g_0+g_x\sm_x+g_y\sm_y+g_z\sm_z \\
        &+ g_{xy}\sm_{xy}+g_{yz}\sm_{yz}+g_{zx}\sm_{zx}+g_{xyz}\sm_{xyz}.
   \end{aligned}
\end{equation}
The number of real degrees of freedom for a multivector is eight, the same as a general $2\times2$ complex matrix. The relation between a multivector and its Pauli matrix representation in $M_2(\mathbb{C})$ is generated by the elements of EQ.~\ref{EQEC: Pauli Mat.}.

Geometric algebras are $\mathbb{Z}_2$-graded, meaning they can be decomposed into two subspaces: A subspace where elements are even multiples of vectors, and a subspace where elements are odd multipes of vectors. Symbolically, these two subspaces are respectively written as $\APS^+$ and $\APS^-$. Scalars are grade-$0$, vectors are grade-$1$, bivectors are grade-$2$, and pseudoscalars (trivectors) are grade-$3$. The algebra can be written as the direct sum of its graded-subspaces,
\begin{equation}\label{EQEC: Graded Subspaces}
    \APS=\bigoplus_{j=0}^3\APS^j = \APS^{0\oplus\cdots\oplus3},
\end{equation}
where $\APS^j$ is the set of grade-$j$ elements. From this concept of grade, the \textit{grade projection} can be defined as
\begin{equation}\label{EQEC: Grade Proj.}
    \left<g\right>_j\in\APS^j,
\end{equation}
satisfying
\begin{equation}\label{EQEC: Grade Proj. Prop.}
    \left<g\right>_{j\oplus k}=\left<g\right>_j+\left<g\right>_k.
\end{equation}
It is often convenient to express multivectors as the sum of their graded subcomponents, and indeed it is necessary for the \textit{Scattering Algebra} of SEC.~\ref{SA}. 

There are three involutory (self-inverse) grade-selecting conjugations inherent to the APS. The first is \textit{grade involution} (in the APS, called \textit{parity conjugation}),
\begin{equation}\label{EQEC: Parity Conjugation}
    g^- = \left<g\right>_0-\left<g\right>_1+\left<g\right>_2-\left<g\right>_3,
\end{equation}
which negates the odd-grade terms. Geometrically, this corresponds to negating all basis directions, which is why this is called parity conjugation in the APS. The second is called \textit{reversion} (in the APS, called \textit{Hermitian conjugation}),
\begin{equation}\label{EQEC: Hermitian Conjugation}
    g^\dagger = \left<g\right>_0+\left<g\right>_1-\left<g\right>_2-\left<g\right>_3,
\end{equation}
which negates bivectors and pseudoscalars. Additionally, for another multivector $h\in\APS$, $(gh)^\dagger=h^\dagger g^\dagger$. The third is \textit{Clifford conjugation}, and is a composition of both parity and Hermitian conjugation,
\begin{equation}\label{EQEC: Clifford Conjugation}
    \widetilde{g}=(g^-)^\dagger=(g^\dagger)^-=\left<g\right>_0-\left<g\right>_1-\left<g\right>_2+\left<g\right>_3,
\end{equation}
which negates vectors and bivectors. Using reversion on a grade-$j$ element, $A\in\APS^j$,
\begin{equation}\label{EQEC: Magnitude}
    |A|=\sqrt{AA^\dagger}
\end{equation}
is the definition of the element's \textit{magnitude}.  

\subsection{$\Spin(3)$ in $\APS$}

The group $\Spin(3)\approx\mathrm{SU}(2)$ decsribes rotations within $3$-dimensional (Euclidean) space. So there should be no surprise that this spin group is natural to the APS,
\begin{equation}\label{EQEC: Spin(3)}
    \Spin(3)=\{R\in\APS^+ \,|\, RR^\dagger=R^\dagger R=1 \},
\end{equation}
where $R\in\Spin(3)$ is called a \textit{rotor}. This group exists within the algebra due to the basis bivectors of EQ.~\ref{EQEC: Biv. Basis}, which form the Lie algebra $\mathfrak{spin}(3)\approx\mathfrak{su}(2)$. In general, rotors are exponentials of an arbitrary bivector $\Omega=|\Omega|\hat{\Omega}\in\APS^2$,
\begin{equation}\label{EQEC: Rotor}
    R = e^{-\frac{1}{2}\Omega}=e^{-\frac{1}{2}|\Omega|\hat{\Omega}}.
\end{equation}
Then, using the \textit{sandwich product},
\begin{equation}\label{EQEC: g -> RgR}
    g\mapsto R gR^\dagger,
\end{equation}
any multivector can be rotated.

\subsection{Spacetime in $\APS$}

The APS is able to describe $(1+3)$-spacetime geometry. This is because the APS is isomorphic to the even subalgebra of the \textit{Spacetime Algebra}, $\APS\approx\eSTA$.  Multivectors which are the sum of scalars and vectors, called \textit{paravectors}, fill the role of the spacetime vector,
\begin{equation}\label{EQEC: Paravector}
    a=\left<a\right>_0+\left<a\right>_1 = a_t+\mathbf{a},
\end{equation}
and are called \textit{spacetime-paravectors}. The product of this spacetime-paravector with its Clifford conjugate $\widetilde{a}$ is
\begin{equation}\label{EQEC: a tilde a}
    \begin{aligned}
        a\widetilde{a}=\widetilde{a}a=aa^-=a^-a=(a_t+\mathbf{a})(a_t-\mathbf{a})&=a_t^2-\mathbf{a}^2 \\
        &= a_t^2-a_x^2-a_y^2-a_z^2,
    \end{aligned}
\end{equation}
where the fact that $a^\dagger=a$ was used. This product is the Lorentz-invariant magnitude of the spacetime-paravector $a$. Given another spacetime-paravector $b\in\APS^{0\oplus1}$, the \textit{spacetime product} is 
\begin{equation}\label{EQEC: Spacetime product}
    a\widetilde{b} = a\sip b+a\sop b.
\end{equation}
The first term is the \textit{spacetime inner product},
\begin{equation}\label{EQEC: ST Inn. Prod.}
    a\sip b=\frac{1}{2}(a\widetilde{b}+b\widetilde{a})=a_tb_t-\mathbf{a}\cdot\mathbf{b},
\end{equation}
while the second term is the \textit{spacetime outer product},
\begin{equation}\label{EQEC: ST Out. Prod.}
    a\sop b = \frac{1}{2}(a\widetilde{b}-b\widetilde{a})=b_t\mathbf{a}-a_t\mathbf{b}-\mathbf{a}\wedge\mathbf{b}.
\end{equation}
There is an opposite parity definition for each spacetime product, obtained by taking the parity conjugation of EQ.~\ref{EQEC: Spacetime product}.

\subsection{$\Spin(1,3)$ in $\APS$}

As the APS is able to describe spacetime geometry, the \textit{Lorentz spin group} exists within the algebra, $\Spin(1,3)\approx\mathrm{SL}(2,\mathbb{C})$:
\begin{equation}\label{EQEC: Lorentz Spin Group}
    \Spin(1,3)=\{\Lambda\in\APS \,|\, \Lambda\widetilde{\Lambda}=\widetilde{\Lambda}\Lambda=1\},
\end{equation}
where $\Lambda\in\Spin(1,3)$ is called a \textit{Lorentz rotor}. The existence of the Lorentz spin group within the APS is due to the basis vectors of EQ.~\ref{EQEC: Basis Vectors} and the basis bivectors of EQ.~\ref{EQEC: Biv. Basis} together forming the Lie algebra $\mathfrak{spin}(1,3)\approx\mathfrak{sl}(2,\mathbb{C})$. In general, Lorentz rotors can be decomposed as
\begin{equation}\label{EQEC: Lorentz Rotor}
    \Lambda = RL,
\end{equation}
where $R$ is a rotor given by EQ.~\ref{EQEC: Rotor} and $L\in\APS^{0\oplus1}$ is a \textit{Lorentz boost}. Explicitly, for a unit direction $\hat{v}\in\APS^1$ and rapidity (hyperbolic angle) $\eta$,
\begin{equation}\label{EQEC: Lorentz Boost}
    L=e^{\frac{1}{2}\eta\hat{v}}
\end{equation}
is a Lorentz boost. Any spacetime-paravector is then Lorentz transformed via the sandwich product,
\begin{equation}
    a\mapsto\Lambda a\Lambda^\dagger.
\end{equation}
Note that this transformation \textit{only} applies to spacetime-paravectors.

\section{Spinor Formalism}\label{SF}

If a particle's spin orientation is given by the unit \textit{spin vector} $\hat{s}\in\APS^1$, then for every possible \textit{reference axis} $\hat{a}\in\APS^1$, there exists a rotor $R_s\in\Spin(3)\subset\APS^+$ such that
\begin{equation}\label{EQSF: Spin Vector}
    \hat{s} = R_s\hat{a}R_s^\dagger.
\end{equation}
This rotor encodes the \textit{spin with respect to} $\hat{a}$, and is the \textit{operatorial Pauli spinor} of \cite{Lounesto2001,Doran2003}. The spin of a particle can be determined using the inner product between the spin vector and the reference axis, $\hat{s}\cdot\hat{a}$. If the product returns $+1$, $\hat{s}\cdot\hat{a}=+1$, then $R_s=1$ is called \textit{spin-up with respect to} $\hat{a}$. Of course, there is a family of $R_s$ which satisfy this condition by leaving $\hat{a}$ invariant. But since such rotors, by definition of invariance, cancel themselves, only $R_s=1$ remains and is therefore the logical choice for spin-up (with respect to $\hat{a}$). If the product returns $-1$, $\hat{s}\cdot\hat{a}=-1$, then by equivalent logic there exists a unit bivector $\hat{B}\in\APS^2$ such that $R_s=\hat{B}$ is called \textit{spin-down with respect to} $\hat{a}$. A general (operatorial) Pauli spinor will be a normalized linear combination of these spin-up and spin-down components. Lastly, if $\hat{p}\in\APS^1$ is the unit vector in the \textit{direction of spatial momentum}, and if the inner product $\hat{s}\cdot\hat{p}=\pm1$ is considered, then $R_s$ can be called the \textit{right/left helicity with respect to} $\hat{a}$. A similar decomposition of this helicity rotor with respect to $\hat{p}$ is possible, and is analogously determined by the afore-written inner product between the spin vector and the unit spatial-momentum.  

\begin{figure}
    \centering
    \includegraphics[width=0.55\linewidth]{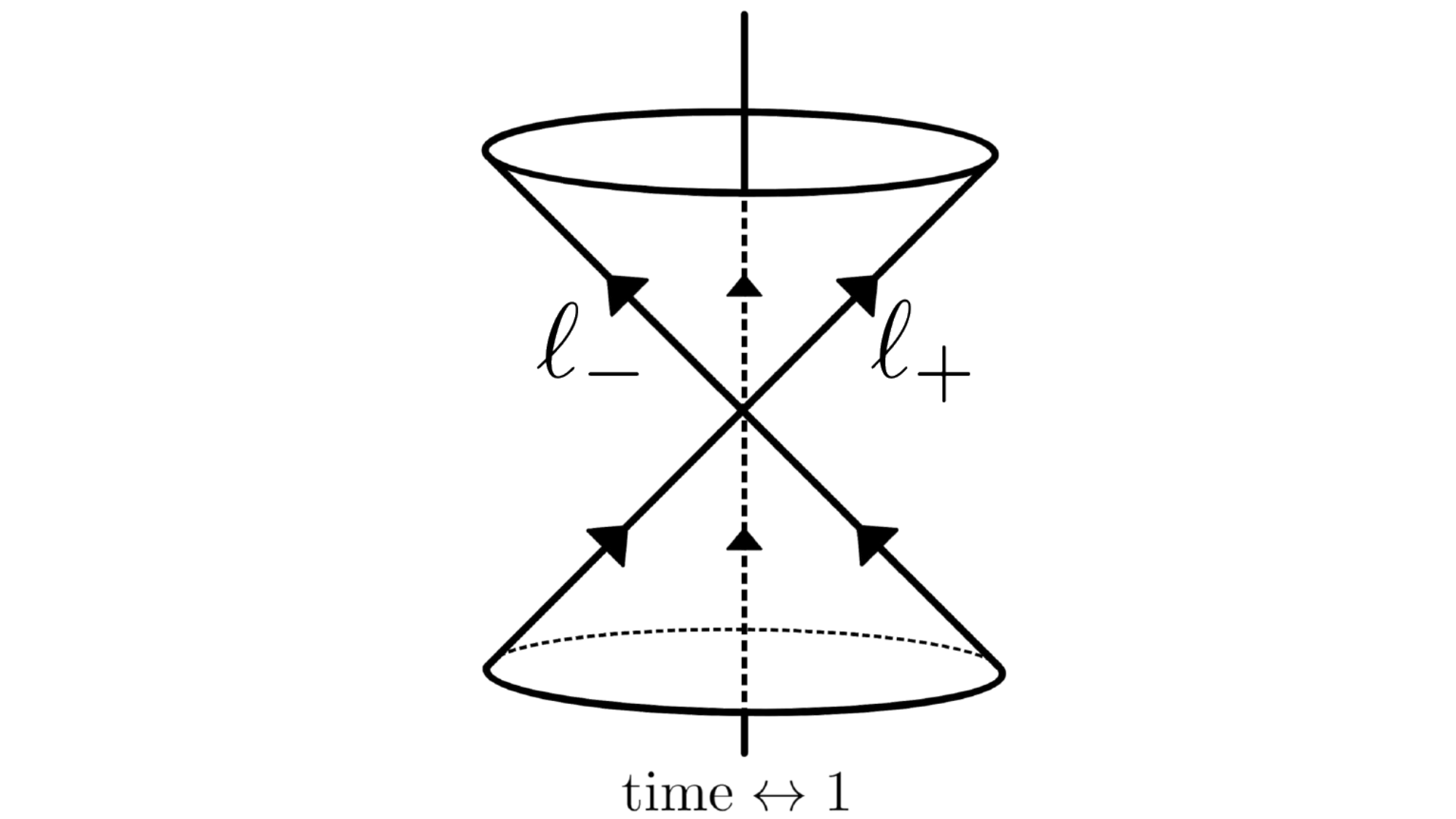}
    \caption{The lightrays $\ell_\pm=(1\pm\hat{a})/2$ shown on the lightcone, partitioning time.}
    \label{FIG:Lightcone}
\end{figure}

It is possible, and indeed relevant for what comes, to introduce the idempotents (projectors)
\begin{equation}\label{EQSF: Idempotents}
    \ell_\pm = \frac{1}{2}(1\pm\hat{a}).
\end{equation}
Geometrically, these describe lightlike trajectories, earning $\ell_\pm$ the title of \textit{lightrays}, in the $\pm\hat{a}$ direction. This is motivated by the fact that within the APS methods for relativity, the unit scalar represents the unit \textit{timelike} direction, while the unit vector represents the unit \textit{spacelike} direction \cite{Hestenes1966,Baylis2004}. Thus EQ.~\ref{EQSF: Idempotents} tells one that time and space are in lockstep, meaning lightlike. The lightrays border the lightcone and partition the timelike direction,
\begin{equation}\label{EQSF: Partition 1}
    1 = \ell_+ + \ell_-.
\end{equation}
This is seen in FIG.~\ref{FIG:Lightcone}. The lightrays $\ell_\pm$ are further called \textit{complementary} because of this partitioning. By this, any multivector $g\in\APS$ can be \textit{lightray-partitioned}, $g = g\ell_++g\ell_-$. These two terms are called \textit{forward-oriented} and \textit{backward-oriented} (light)\textit{ray} spinors,
\begin{equation}\label{EQSF: Weyl spinors}
    \al_{\Fo} = g\ell_+\quad\text{and}\quad\be_{\Ba} = g\ell_-.
\end{equation}
For simplicity, and if differentiating between the two is needed, the first kind will be called an F-ray spinor and the second kind a B-ray spinor. Like the lightrays which form them, these ray spinors are called complementary since they add to the multivector $g$. The F-ray spinor lives in the \textit{minimal left ideal} $\APS \ell_+$, while its complementary B-ray spinor lives in the \textit{mutually independent} minimal left ideal $\APS \ell_-$. The structure of these ray spinors, to be given near this section's end, is equivalent to the \textit{Vaz-da Rocha formalism} in \cite{Vaz2019}, to the \textit{Penrose-Rindler formalism} in \cite{Penrose1986}, and to the \textit{two-spinor calculus} in \cite{Doran2003}. In general, for each reference axis $\hat{a}$ there exists a perpendicular unit vector $\hat{b}\in\APS^1$ satisfying $\hat{a}\cdot\hat{b}=0$.  It arises in the expansion of the ray spinors,
\begin{equation}\label{EQSF: L/R Spinors}
    \begin{aligned}
        g\ell_++g\ell_- &= \al_\Fo+\be_\Ba \\
        &= \al_\uparrow \ell_+ + \al_\downarrow \hat{b}\ell_+ + \be_\uparrow \ell_-+\be_\downarrow\hat{b}\ell_-,
    \end{aligned}
\end{equation}
for \textit{spin-components} $\al_{\uparrow/\downarrow},\be_{\uparrow/\downarrow}\in\mathrm{Center[\APS]=\APS^{0\oplus 3}\approx \mathbb{C}}$ with respect to $\hat{a}$. However, this $\hat{b}$ is not wholly new. It it related to the spin-down rotor (with respect to $\hat{a}$) through
\begin{equation}\label{EQSF: B and b}
    \hat{B}\ell_\pm = \pm\hat{b}\ell_\pm.
\end{equation}
This implies that, from EQ.~\ref{EQSF: L/R Spinors} and invoking $g=R_s$, the $\al_\uparrow \ell_+$ and $\be_\uparrow \ell_-$ terms are the spin-up components of the Pauli spinor $R_s$ (with respect to $\hat{a}$) projected onto each of the lightrays, $\ell_\pm$. Likewise, the $\al_\downarrow \hat{b}\ell_+$ and $\be_\downarrow\hat{b}\ell_-$ terms are the spin-down components of this Pauli spinor $R_s$ projected onto each of the lightrays, $\ell_\pm$. 

Thus far, all equations have been presented in general forms based upon the spin-measurement direction $\hat{a}$. In physics it is standard practice to set $\hat{a}=\sm_z$. This means that one makes reference along the $z$-direction. Doing so implies that the spin-down component with respect to $\sm_z$ is $\sm_{xz}$, and that EQ.~\ref{EQSF: B and b} becomes $\sm_{xz}\ell_\pm = \pm\sm_x \ell_\pm$. In this case, it is easier to connect with traditional matrix formalisms by mapping to the \textit{Pauli representation} in $\mathrm{M}_2(\mathbb{C})$, the algebra of $2\times2$ complex-entry matrices, as detailed in SEC.~\ref{A0}. However, aside from the dictionary in SEC.~\ref{SA}, this paper will continue to work coordinate-free with $\hat{a}$.

\subsection{Lorentz Spinors}

The spacetime-paravector (scalar plus vector) $p=E/c+\mathbf{p}\in\APS^{0\oplus 1}$ is called the \textit{spacetime momentum} (for energy $E$ and speed of light $c$), and can be constructed from the Lorentz transformation of a rest momentum,
\begin{equation}\label{EQSF: p = RLLR}
    p = \Lambda mc \Lambda^\dagger,
\end{equation}
where  generally $\Lambda=RL\in\Spin(1,3)$ is a composition of a rotor $R\in\Spin(3)$ and a \textit{Lorentz boost} $L\in\APS^{0\oplus 1}$ satisfying $L\widetilde{L}=\widetilde{L}L=1$. This implies $\Lambda\widetilde{\Lambda}=\widetilde{\Lambda}\Lambda=1$. As will be shown near the end of SEC.~\ref{SA}, the method of the Constructive Standard Model (CSM) conventionally\footnote{The conventional method of the CSM actually corresponds to $\hat{a}=\sm_z$.} uses $\Lambda = R_pL_a$, where $L_a$ is a boost in the $\hat{a}$ direction and $R_p$ is the rotor rotating $\hat{a}$ into $\hat{p}$, the unit spatial-momentum direction. However, it is possible to simply boost directly into the spatial-momentum direction with rapidity $\eta=\tanh^{-1}{\beta_\mathrm{rv}}$ (where $\beta_\mathrm{rv}=|\mathbf{v}|/c$ is relative velocity), 
\begin{equation}\label{EQSF: L_p}
    L_p = e^{\frac{1}{2}\eta\hat{p}}=\cosh{\frac{\eta}{2}}+\hat{p}\sinh{\frac{\eta}{2}},
\end{equation}
as this boost satisfies
\begin{equation}\label{EQSF: p = LpLp}
    p = L_p mcL_p^\dagger.
\end{equation}
The proof of equality between EQ.~\ref{EQSF: p = RLLR} and EQ.~\ref{EQSF: p = LpLp} is simple:
\begin{align*}
    R_pL_aL_a^\dagger R_p^\dagger &= R_pL_aR_p^\dagger R_pL_a^\dagger R_p^\dagger \\
    &= L_pL_p^\dagger,
\end{align*}
where both $R_p^\dagger R_p = 1$ and $R_p L_a R_p^\dagger=L_p$ were used. Because $R_p$ by definition satisfies $\hat{p}=R_p \hat{a} R_p^\dagger$, one might think this rotor can be called the right-helicity with respect to $\hat{a}$. But, it is important to realize that the rotors within a Lorentz transformation are \textit{not} the rotors involved in the definitions of spin in EQ.~\ref{EQSF: Spin Vector}. Mathematically, $R_p\neq R_s$. This means that $R_p$ is not actually describing the spin-orientation of the particle, so calling it right-helicity should be avoided. 

It will be convenient to rewrite the spacetime momentum as either
\begin{equation}\label{EQSF: p = ll}
    p = \lambda \lambda^\dagger\quad\text{or}\quad p = \lambda_p\lambda_p^\dagger,
\end{equation}
where $\lambda = \sqrt{mc}R_pL_a$ and $\lambda_p = \sqrt{mc}L_p$ are respectively called \textit{chiral} and \textit{achiral} \textit{Lorentz spinors}. A discussion on chirality is given in SEC.~\ref{A0.5}. These Lorentz spinors satisfy the \textit{mass condition},
\begin{equation}\label{EQSF: Mass Condition}
    \lambda\widetilde{\lambda}=\widetilde{\lambda}\lambda = 
     \lambda^-\lambda^\dagger = \lambda^\dagger \lambda^- = 
    \lambda_p\widetilde{\lambda}_p=\widetilde{\lambda}_p\lambda_p = mc,
\end{equation}
which is a logical consequence of the earlier normalization condition $\Lambda\widetilde{\Lambda}=\widetilde{\Lambda}\Lambda=1$. When lightray-partitioned using the lightrays from EQ.~\ref{EQSF: Idempotents}, the chiral Lorentz spinor gives
\begin{equation}\label{EQSF: l lightray-part.}
    \begin{aligned}
    \lambda \ell_++\lambda \ell_- &= e^{\frac{\eta}{2}}\sqrt{mc}R_p \ell_+ + e^{-\frac{\eta}{2}}\sqrt{mc}R_p\ell_- \\
    &= \sqrt{\frac{1}{c}E+|\mathbf{p}|}(\xi_{+,\uparrow}\ell_+ + \xi_{+,\downarrow}\hat{b}\ell_+) \\ 
    &+ \sqrt{\frac{1}{c}E-|\mathbf{p}|}(\xi_{-,\uparrow}\ell_- + \xi_{-,\downarrow}\hat{b}\ell_-).
\end{aligned}
\end{equation}
The positive square-root eigenvalue states that the spatial-momentum (before $R_p$ rotates it to $\hat{p}$) is \textit{positively-aligned} with the lightray $\ell_+$, while the negative square-root eigenvalue states that the spatial-momentum (before $R_p$ rotates it to $\hat{p}$) is \textit{negatively-aligned} with the lightray $\ell_-$.  Meanwhile, the factors in parentheses are simply the rotor $R_p$ projected on each lightray. While the coefficients $\xi_{\pm,\uparrow/\downarrow}$ have arrows indicating that the rotor has been decomposed into its spin-components with respect to the reference axis $\hat{a}$, they do not give the spin of the particle with spacetime momentum $p$. It is just a basis decomposition, stating what components are aligned with the spin-up and spin-down elements of the spin defined by EQ.~\ref{EQSF: Spin Vector}. For consistency, these coefficients are related to EQ.~\ref{EQSF: L/R Spinors} by $\al_{\uparrow/\downarrow}=\sqrt{E/c+|\mathbf{p}|}\xi_{+,\uparrow/\downarrow}$ and $\be_{\uparrow/\downarrow}=\sqrt{E/c-|\mathbf{p}|}\xi_{-,\uparrow/\downarrow}$. Continuing to apply the lightray-partitioning to the achiral Lorentz spinor $\lambda_p$ using the lightrays $\ell_\pm$ gives only
\begin{equation}\label{EQSF: l_p lightray-part}
    \lambda_p \ell_+ + \lambda_p \ell_- = \sqrt{mc}(\zeta_{+,\uparrow}\ell_++\zeta_{+,\downarrow}\hat{b}\ell_+) + \sqrt{mc}(\zeta_{-,\uparrow}\ell_-+\zeta_{-,\downarrow}\hat{b}\ell_-),
\end{equation}
which is arguably uninteresting since it only projects the spinor upon the lightrays $\ell_\pm$, as has already been demonstrated many times. Moreover, the square-root eigenvalues hold no information regarding lightray-alignment. It is more insightful to instead define another pair of lightrays,
\begin{equation}\label{EQSF: l_p lightrays}
    p_\pm = \frac{1}{2}(1\pm\hat{p}).
\end{equation}
Applying these gives
\begin{equation}\label{EQSF: l_p lightray-part 2}
   \lambda_p p_+ + \lambda_p p_- = \sqrt{\frac{1}{c}E+|\mathbf{p}|}p_+ + \sqrt{\frac{1}{c}E-|\mathbf{p}|}p_-. 
\end{equation}
As with $\ell_\pm$ and $\lambda$, the positive square-root eigenvalue states that the spatial-momentum is positively-aligned with the lightray $p_+$, while the negative square-root eigenvalue states that the spatial-momentum is negatively-aligned with the lightray $p_-$. Setting $\hat{a}=\hat{p}$ would imply that $R_p=1$ in $\lambda$, thereby equating EQ.~\ref{EQSF: l lightray-part.} and EQ.~\ref{EQSF: l_p lightray-part 2}. Thus, for any achiral Lorentz spinor, the lightrays $p_\pm$ are used to obtain the alignment eigenvalues. 

It might seem like an unnecessary tangent to explain the lightray-partitioned geometry of these Lorentz spinors. But this section will be a helpful tool for Geometric Algebraists trying to learn the standard approach to the Constructive Standard Model (CSM) which uses matrices with entries equivalent to the alignment-eigenvalues and spin-components of this subsection. This section will \textit{also} be a helpful tool for traditional physicists trying to learn the Geometric Algebra approach to the CSM. The information of this section will be of further relevance in the discussions near the end of SEC.~\ref{SA}.

\subsection{Ray Spinor Structure}

The product which forms the foundation of the Scattering Algebra introduced in SEC.~\ref{SA} will be seen to satisfy the \textit{ray spinor structure}. As such, it is important to describe it. Here, the lightrays $\ell_\pm$ are used. But there would be an identical structure for \textit{any} complementary pair of lightrays.

\textit{Left/right ideals} are subsets of an algebra wherein an element of the ideal, whenever multiplied on the left/right by an element of the algebra, stays inside the ideal. Minimal ideals contain no further ideals. As mentioned in the section introducing the lightray-partitioning of a multivector, the Algebra of Physical Space can be partitioned into two mutually-independent minimal left ideals, $\APS \ell_+$ and $\APS \ell_-$. At the risk of a redundant statement, any multivector $g\in\APS$ can be lightray-partitioned into two complementary ray spinors,
\begin{equation}\label{EQSF: LW and RW}
    g = g\ell_++g\ell_- = \al_{\Fo}+\be_{\Ba}.
\end{equation}
F-ray and B-ray spinors swap lightray-orientation via parity conjugation,
\begin{equation}\label{EQSF: LW Parity}
    \al_{\Ba} = \al_{\Fo}^-=g^-\ell_-\in\APS \ell_-
\end{equation}
and 
\begin{equation}\label{EQSF: RW Parity}
    \be_{\Fo} = \be_{\Ba}^-=g^-\ell_+\in\APS \ell_+,
\end{equation}
which also leads to swapping ideals. Now, ray spinor structure also involves the minimal \textit{right} ideals, $\ell_+\APS$ and $\ell_-\APS$, which are algebraically dual to the left ideals. Ray spinors are then mapped to these right ideals via
\begin{equation}\label{EQSF: LW to DLW}
    \al^{\Fo}=\hat{b}\widetilde{\al}_{\Fo}=\ell_+\hat{b}\widetilde{g}\in \ell_+\APS
\end{equation}
and
\begin{equation}\label{EQSF: RW to DRW}
    \be^{\Ba}=\hat{b}\widetilde{\be}_{\Ba}=\ell_-\hat{b}\widetilde{g}\in \ell_-\APS,
\end{equation}
which are each called a \textit{dual} ray spinor. Dual ray spinors are just ray spinors, first Clifford conjugated, then left-multiplied by $\hat{b}$ (the vector relating to spin-down, from EQ.~\ref{EQSF: B and b}). As a whole, this swaps spin-components, inserts a relative minus sign on one of the spin-components, and maps from the original ray spinor's ideal to its dual:
\begin{equation}\label{EQSF: spinor to dual}
    \psi_{\Fo/\Ba}=\psi_{\pm,\uparrow}\ell_{\pm}+\psi_{\pm,\downarrow}\hat{b}\ell_\pm\,\,\mapsto\,\,\psi^{\Fo/\Ba}=\ell_\pm\hat{b}\psi_{\pm,\uparrow}-\ell_\pm\psi_{\pm,\downarrow},
\end{equation}
where again $\psi_{\pm,\uparrow/\downarrow}\in\mathrm{Center}[\APS]=\APS^{0\oplus3}\approx\mathbb{C}$.
\begin{figure}
    \centering
    \includegraphics[width=0.5\linewidth]{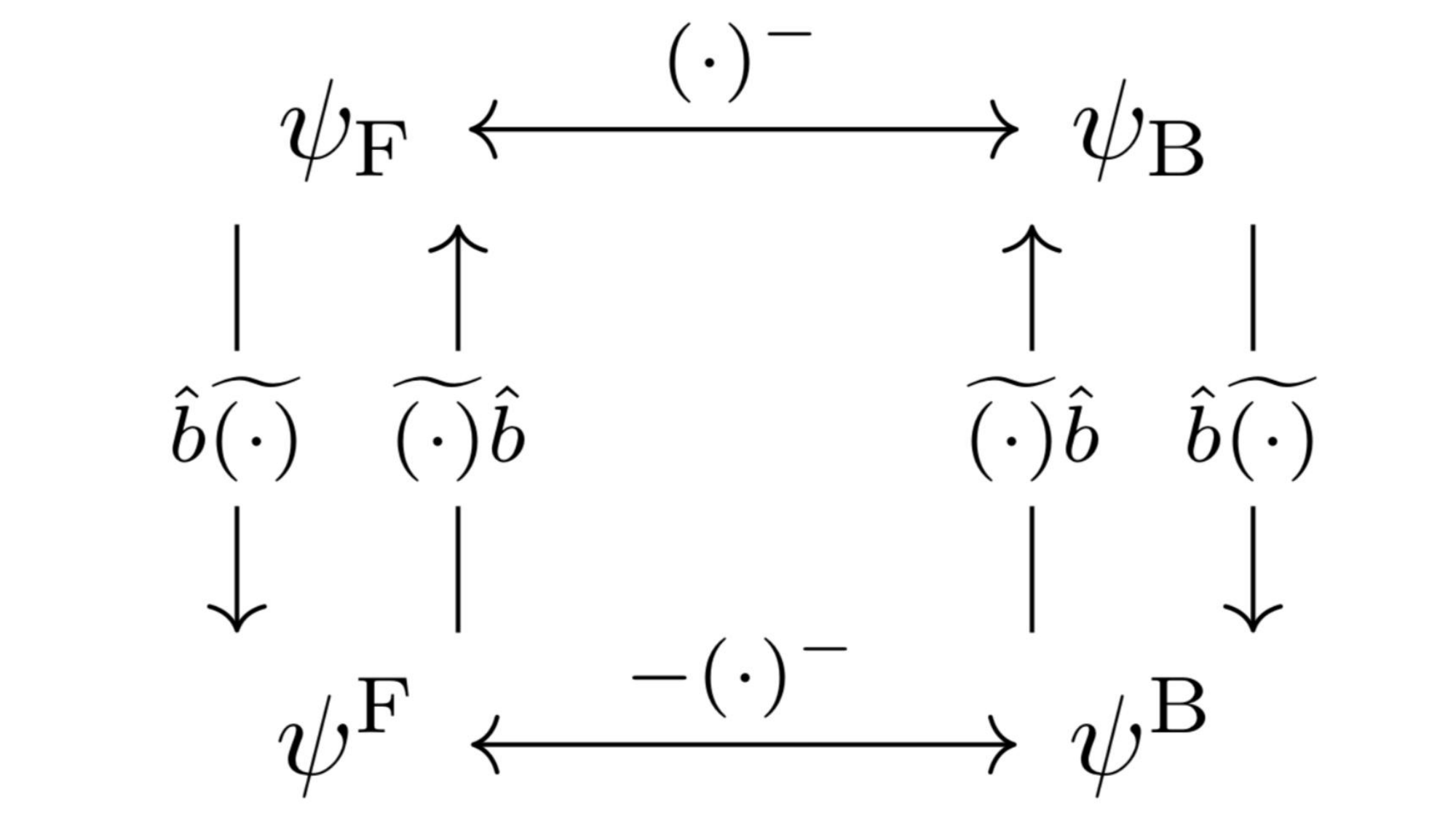}
    \caption{A directed-graph summary of the (light)ray spinor structure in the Algebra of Physical Space.}
    \label{FIG:Weyl Structure}
\end{figure}
This structure is summarized by the directed graph in FIG.~\ref{FIG:Weyl Structure}.
The notation was defined in such a way as to encode for a \textit{symplectic} Lorentz-invariant \textit{ray spinor product}. Supposing $\psi_{\Fo/\Ba}$ and $\phi_{\Fo/\Ba}$ are non-complementary ray spinors, then the ray spinor product is
\begin{equation}\label{EQSF: Weyl Product}
    \psi^{\Fo/\Ba}\phi_{\Fo/\Ba}=\hat{b}\widetilde{\psi}_{\Fo/\Ba}\phi_{\Fo/\Ba}.
\end{equation}
It can be easily verified that if the spinors are equal, $\psi_{\Fo/\Ba}=\phi_{\Fo/\Ba}$, then the product vanishes, which is equivalent to $\psi^{\Fo/\Ba}\phi_{\Fo/\Ba}=-\phi^{\Fo/\Ba}\psi_{\Fo/\Ba}$. A Lorentz-invariant ray spinor product also exists in a \textit{non-symplectic} form between an F-ray spinor and B-ray spinor. Using $\al_{\Fo/\Ba}$ and $\be_{\Ba/\Fo}$:
\begin{equation}\label{EQSF: Weyl Product 2}
    \al^{\Fo/\Ba}\be_{\Ba/\Fo}\quad\text{or}\quad\be^{\Ba/\Fo}\al_{\Fo/\Ba}.
\end{equation}
It can be easily proven that $\left<\hat{b}\al^{\Fo/\Ba}\be_{\Ba/\Fo}\right>_{0\oplus3}=\left<\be^{\Ba/\Fo}\al_{\Fo/\Ba}\hat{b}\right>_{0\oplus3}$, which precludes this product from being symplectic.

In summary, ray spinors exist as the partitioning, or projecting, of multivectors onto both, or one, of a pair of lightrays. Complementary F-ray and B-ray spinors swap lightray-orientation under parity conjugation, which is identical to saying that F-ray and B-ray spinors swap complementary ideals under parity conjugation.  Dual ray spinors are formed via the Clifford conjugate followed by a left-multiplication by $\hat{b}$. Together, ray spinors and their duals with identical lightray-orientation form a (symplectic) Lorentz-invariant product. And there is a (non-symplectic) Lorentz-invariant product between ray spinors and dual ray spinors which don't share lightray-orientation.

\section{Scattering Algebra}\label{SA}

The \textit{Scattering Algebra} (SA) is a tool\footnote{The Scattering Algebra \textit{does not} satisfy the definitions of an algebra and the use of "algebra" is purely nominal.} within the Algebra of Physical Space (APS) that allows for the calculation of constructive scattering amplitudes---and moreover connects helicity methods to the APS---and it is built upon Lorentz spinors. It is important to note that while there exist chiral and achiral Lorentz spinors, the results for amplitude squares and interaction terms are the same regardless of which is used. Therefore, without loss of generality, this section will take $\lambda$ to be a \textit{chirality-agnostic} Lorentz spinor satisfying EQ.~\ref{EQSF: p = ll}. Furthermore, the \textit{natural-unit convention} will be adopted ($c=1$). The SA is based off the Lorentz-invariant product, herein called the \textit{Lorentz product},
\begin{equation}\label{EQSA: LI Product}
    \hat{B}\widetilde{\lambda}_j\lambda_k,
\end{equation}
where $j$ and $k$ are particle labels, and $\hat{B}$ is the spin-down rotor with respect to the reference axis $\hat{a}$. Using the lightrays $\ell_\pm$ defined by EQ.~\ref{EQSF: Idempotents}, the Lorentz product decomposes into the symplectic and non-symplectic ray spinor products,
\begin{equation}\label{EQSA: LC Product Decomposition}
    \hat{B}\widetilde{\lambda}_j\lambda_k = \be^\Ba_j\be_\Ba^k-\al^\Fo_j\al_\Fo^k+\be^\Ba_j\al_\Fo^k-\al^\Fo_j\be_\Ba^k.
\end{equation}
Here $j$ and $k$, both superscript and subscript, continue to serve as particle labels. Regardless of the chirality of the Lorentz spinors, the above Lorentz products describe left-chiral \textit{interactions}. There is also a Lorentz product describing right-chiral interactions, obtained using the ray spinor structure\footnote{Using $\lambda_k\mapsto\lambda_k^-$ and $\hat{B}\widetilde{\lambda}_j\mapsto-(\hat{B}\widetilde{\lambda}_j)^-$, which is the structure summarized in FIG.~\ref{FIG:Weyl Structure}.},
\begin{equation}\label{EQSA: RC LI Product}
    \begin{aligned}
        -\left(\hat{B}\widetilde{\lambda}_j\lambda_k\right)^- &= \hat{B}^\dagger\lambda^\dagger_j\lambda_k^- \\
        &= \be^\Fo_j\be_\Fo^k-\al^\Ba_j\al_\Ba^k+\be^\Fo_j\al_\Ba^k-\al^\Ba_j\be_\Fo^k.
    \end{aligned}
\end{equation}
 Of course, if the particles are the same ($j=k$), then the symplectic contributions to both EQ.~\ref{EQSA: LC Product Decomposition} and EQ.~\ref{EQSA: RC LI Product} vanish. Likewise, the non-symplectic terms remain and return
 \begin{equation}\label{EQSA: Mass identity}
     \begin{aligned}
         \hat{B}\widetilde{\lambda}_j\lambda_j = -\hat{B}^\dagger\lambda_j^\dagger\lambda_j^- &= \be^\Ba_j\al_\Fo^j-\al^\Fo_j\be_\Ba^j \\
         &= \al^\Ba_j\be_\Fo^j-\be^\Fo_j\al_\Ba^j = \hat{B}m_j,
     \end{aligned}
 \end{equation}
which is equivalent to the mass identities\footnote{Equations \textit{A27} and \textit{A28}.} in the appendix of \cite{Christensen2018}. While this is indeed a trivial calculation given the fact it is a restatement of the mass condition in EQ.~\ref{EQSF: Mass Condition}, the non-symplectic ray spinor products show another way of looking at the mass identity of EQ.~\ref{EQSA: Mass identity}: The cross-interaction of F-ray and B-ray spinors which exist on the lightcone. 

However, each Lorentz product is not yet in the SA. The \textit{Scattering Algebra} is taken to be an abstract space wherein Lorentz products are each placed in their own subspace \cite{Christensen2018}. This is conceptually compatible with the APS approach, but it turns out that the SA exists fully \textit{inside} the APS. If $g\in\APS$ is some multivector, then its \textit{SA bracket} is given by
\begin{equation}\label{EQSA: SA Bracket}
    \upSAb{g}{\jj}{\kk} = \ell_\jj g\ell_\kk + \hat{b}\ell_\jj g\ell_\kk\hat{b},
\end{equation}
where $\jj,\kk\in\{+,-\}$ are indices over lightray-orientation. The Hermitian conjugate of the SA bracket is then
\begin{equation}\label{EQSA: SA Bracket HC}
    \left(\upSAb{g}{\jj}{\kk}\right)^\dagg= \downSAb{g^\dagg}{\jj}{\kk} = \ell_\kk g^\dagg\ell_\jj + \hat{b}\ell_\kk g^\dagg\ell_\jj\hat{b}.
\end{equation}
The raising and lowering of indices for SA brackets is simply a visual convention to indicate summation. Clearly, the form of the two above equations demonstrates there is no algebraic difference between upper and lower. Geometrically, the indices select out magnitudes of components of the multivector which have different lightray-alignments. The rightmost index $\kk$ selects out a specific ray spinor from the multivector, while the leftmost index $\jj$ selects the component of this ray spinor that is either parallel or perpendicular to the lightray $\ell_\kk$. Algebraically, the indices select out values corresponding to the elements of the matrix representation induced by $\hat{a}$. The construction of EQ.~\ref{EQSA: SA Bracket} ensures equivalence between the traditional CSM approach and the APS approach. It can be easily proven that for multivectors $g_j\in\APS$, using the Einstein summation convention,
\begin{equation}\label{EQSA: g g g g}
    \left[g_1\right]^{\mathrm{AB}}\left[g_2\right]_{\mathrm{BC}}\left[g_3\right]^\mathrm{CD}\dots = 2\left<g_1g_2g_3\dots\right>_{0\oplus 3},
\end{equation}
where $\cproj{\cdot}$ is the projection to the grade-$0$ and grade-$3$ subspaces, called the \textit{central projection}, and the final index of summation is a lowered $\mathrm{A}$. This use of the central projection to simplify squared amplitudes is identical to the standard trace methods of \cite{Christensen2020,2Christensen2024,3Christensen2024,4Christensen2024}.
To obtain the SA of the traditional CSM, the Lorentz products are placed into their own SA brackets:
\begin{equation}\label{EQSA: Mapping to SA}
    \begin{aligned}
        \hat{B}\widetilde{\lambda}_j\lambda_k &\longmapsto \left[\hat{B}\widetilde{\lambda}_j\lambda_k\right]^{\jj\kk} \\
        \hat{B}^\dagger\lambda^\dagger_j\lambda_k^- &\longmapsto \left[\hat{B}^\dagger\lambda^\dagger_j\lambda_k^-\right]^{\jj\kk}.
    \end{aligned}
\end{equation}
This is a map from the APS to the SA (also within the APS). While a meaningful tool, it will be seen that the indices are redundant. The brackets of Lorentz products have indices (and particle labels) that are antisymmetric under particle-exchange,
\begin{equation}\label{EQSA: Antisymmetric Particle Exchange}
     \begin{aligned}
        \left[\hat{B}\widetilde{\lambda}_j\lambda_k\right]^{\jj\kk} &=  -\left[\hat{B}\widetilde{\lambda}_k\lambda_j\right]^{\kk\jj} \\
        \left[\hat{B}^\dagger\lambda^\dagger_j\lambda_k^-\right]^{\jj\kk} &= -\left[\hat{B}^\dagger\lambda^\dagger_k\lambda_j^-\right]^{\kk\jj}. 
    \end{aligned}
\end{equation}
These identities are crucial for correctly simplifying more complicated squared scattering amplitudes. The \textit{Hermitian conjugate} of both brackets are 
\begin{equation}\label{EQSA: Herm. Conj.}
    \begin{aligned}
        \left(\left[\hat{B}\widetilde{\lambda}_j\lambda_k\right]^{\jj\kk}\right)^\dagger &= \left[\lambda^\dagger_k\lambda_j^-\hat{B}^\dagger\right]_{\kk\jj} \\
        \left(\left[\hat{B}^\dagger\lambda^\dagger_j\lambda_k^-\right]^{\jj\kk}\right)^\dagger &= \left[\widetilde{\lambda}_k\lambda_j\hat{B}\right]_{\kk\jj}. 
    \end{aligned}
\end{equation}
Squared constructive amplitudes deal with the sum over indices of an amplitude times its Hermitian conjugate. Successfully completing such a square returns twice the central projection, seen in EQ.~\ref{EQSA: g g g g}, which can be easily simplified using the well-known properties covered in the fourth chapter of \cite{Doran2003}. An example, the square of the left-chiral Lorentz product, is then calculated to be
\begin{equation}\label{EQSA: LC LP Square}
    \begin{aligned}
        \left[\hat{B}\widetilde{\lambda}_j\lambda_k\right]^{\jj\kk}\left[\lambda^\dagger_k\lambda_j^-\hat{B}^\dagger\right]_{\kk\jj} &= 2\left<\hat{B}\widetilde{\lambda}_j\lambda_k\lambda_k^\dagger\lambda_j^-\hat{B}^\dagger\right>_{0\oplus 3} \\
    &= 2\left<p^-_jp_k\right>_{0\oplus 3} = 2p_j\sip p_k,
    \end{aligned}
\end{equation}
where $\lambda^-_j\widetilde{\lambda}_j=p^-_j$ and $\lambda_k\lambda_k^\dagger=p_k$ were invoked using EQ.~\ref{EQSF: p = ll}, and the final term $p_j\sip p_k$ is the (paravector) spacetime inner product defined in EQ.~\ref{EQEC: ST Inn. Prod.}.  

Before continuing on, there is one last bracket in the SA that needs discussing: That of \textit{momentum-insertion}. A momentum-insertion product is given by
\begin{equation}\label{EQSA: SA Mom-Ins}
    \left[\hat{B}\lambda_j^\dagger p_l^-\lambda_k\right]^{\jj\kk}
    \quad\text{or}\quad
    \left[\hat{B}^\dagger\widetilde{\lambda}_j p_l\lambda_k^-\right]^{\jj\kk}.
\end{equation}
The indices (and particle labels) are antisymmetric under particle-exchange,
\begin{equation}\label{EQSA: Particle Exchange Mom-Ins}
    \left[\hat{B}\lambda_j^\dagger p_l^-\lambda_k\right]^{\jj\kk} = -\left[\hat{B}\widetilde{\lambda}_k p_l\lambda_j^-\right]^{\kk\jj}=\left[\hat{B}^\dagger\widetilde{\lambda}_k p_l\lambda_j^-\right]^{\kk\jj},
\end{equation}
however it is not as straightforward. Indeed there was a negative sign as a result of particle-exchange, but there was also\footnote{This implies the involvement of the ray spinor structure.} an overall parity conjugation. Thus the two different momentum-insertion products are equal within the SA. Like the antisymmetric identity of EQ.~\ref{EQSA: Antisymmetric Particle Exchange}, the above identity will likewise be necessary in the simplification of certain squared amplitudes.  

\subsection{Initial Dictionary between Traditional and Geometric Formalism}

The chiral Lorentz spinors of the APS correspond directly with the \textit{spin spinors} of the Constructive Standard Model (CSM). The chiral Lorentz spinor $\lambda$ corresponds to the \textit{right-angle} (\textit{spin}) \textit{spinor}, $|\mathbf{j}\rangle^{\ \jj}_\al$, where $\jj\in\{1,2\}$ is a \textit{spin-index} (row-index) and $\al\in\{1,2\}$ is a \textit{Lorentz-index} (column-index). Specifically, when the reference axis $\hat{a}$ is chosen to be $\sigma_z$, the latter is the matrix representation of the former:
\begin{equation}\label{EQSA: lambda and rangle}
    \lambda = \sqrt{m}R_pL_z\equiv\begin{bmatrix}
        \sqrt{E+|\mathbf{p}|}\cos{\frac{\theta}{2}} & -\sqrt{E-|\mathbf{p}|}\sin{\frac{\theta}{2}}e^{-i\omega} \\
        \sqrt{E+|\mathbf{p}|}\sin{\frac{\theta}{2}}e^{i\omega} & \sqrt{E-|\mathbf{p}|}\cos{\frac{\theta}{2}}
    \end{bmatrix}\leftrightarrow|\mathbf{j}\rangle_\al^{\ \jj}.
\end{equation}
Here, $\theta$ is the polar angle and $\omega$ is the azimuthal angle determined by $R_p$, which in this case rotates from $\sigma_z$ to the spatial-momentum direction $\hat{p}$. Notice that the traditional CSM notation for an angle spinor\footnote{All spin spinors take on \textit{scalar values} for specific index choices, but are represented by a matrix--as seen in EQ.~\ref{EQSA: lambda and rangle} and EQ.~\ref{EQSA: lambdadagg and langle}. These matrices are the representations of their corresponding Lorentz spinors.} is not a ket, despite having similar notation. The above correspondence is fully equivalent to the lightray-partitioning of EQ.~\ref{EQSF: l lightray-part.}, so the geometric interpretation of the matrix elements is the same. The Hermitian conjugate (reverse) of this Lorentz spinor, $\lambda^\dagger$, corresponds to the \textit{left-square} (\textit{spin}) \textit{spinor}, $\lbrack\mathbf{j}|_{\dot{\be}\jj}$. The latter is the \textit{transpose} of the matrix representation of the former:
\begin{equation}\label{EQSA: lambdadagg and langle}
    \begin{aligned}
        \lambda^\dagger = \sqrt{m}L_zR_p^\dagger&\equiv
        \begin{bmatrix}
        \sqrt{E+|\mathbf{p}|}\cos{\frac{\theta}{2}} & \sqrt{E+|\mathbf{p}|}\sin{\frac{\theta}{2}}e^{-i\omega} \\
        -\sqrt{E-|\mathbf{p}|}\sin{\frac{\theta}{2}}e^{i\omega} & \sqrt{E-|\mathbf{p}|}\cos{\frac{\theta}{2}}
    \end{bmatrix}\\
        &\leftrightarrow\left(\lbrack\mathbf{j}|_{\dot{\be}\jj}\right)^T.
    \end{aligned}
\end{equation}
It is seen that,
\begin{equation}\label{EQSA: p and p al be}
    \lambda\lambda^\dagger = p \quad\leftrightarrow\quad |\mathbf{j}\rangle^{\ \jj}_\al\lbrack\mathbf{j}|_{\dot{\be}\jj}=p_{\al\dot{\be}},
\end{equation}
which is the \textit{momentum product} of \cite{Christensen2024,Christensen2018}, where $p_{\al\dot{\be}}$ is the matrix representation of the spacetime momentum $p$. The further correspondences between the standard CSM spin spinors and APS-CSM Lorentz spinors are: That the \textit{left-angle} (\textit{spin}) \textit{spinor} $\langle\mathbf{j}|^{\al\jj}$ is the transpose of the matrix representation of $\sm_{xz}\widetilde{\lambda}$, and the \textit{right-square} (\textit{spin}) \textit{spinor} $|\mathbf{j}\rbrack^{\dot{\be}}_{\ \jj}$ is the matrix representation of $\lambda^-\sm_{zx}$. Therefore,
\begin{equation}\label{EQSA: p and p al be 2}
    \lambda^-\widetilde{\lambda} = p^- \quad\leftrightarrow\quad |\mathbf{j}\rbrack^{\dot{\be}}_{\ \jj}\langle\mathbf{j}|^{\al\jj}=p^{\dot{\be}\al}
\end{equation}
is the second momentum product of \cite{Christensen2024,Christensen2018}, where $p^{\dot{\be}\al}$ is the matrix representation of the (parity conjugated) spacetime momentum $p^-$. The \textit{left-chiral spinor product} of the standard CSM (with raised spin-indices) is fully equivalent to the left-chiral Lorentz product,
\begin{equation}\label{EQSA: LC products}
    \sm_{xz}\widetilde{\lambda}_j\lambda_k \quad\leftrightarrow\quad \langle\mathbf{j}|^{\al\jj}|\mathbf{k}\rangle_\al^{\ \kk}.
\end{equation}
The \textit{right-chiral spinor product} of the standard CSM naturally has lowered spin-indices. In the standard method for the CSM, all components of an amplitude must have identically-positioned spin-indices, which means that the right-chiral spinor product must have its spin-indices raised by an object called the \textit{epsilon tensor}, $\epsilon^{\jj\kk}$. This object is simply the matrix representation for $\sm_{xz}$. Therefore the right-chiral spinor product of the standard CSM (with raised spin-indices) is fully equivalent to the right-chiral Lorentz product,
\begin{equation}\label{EQSA: RC products}
    \sm_{zx}\lambda^\dagger_j\lambda_k^- \quad\leftrightarrow\quad \lbrack\mathbf{j}|_{\dot{\be}\mm}\epsilon^{\mm\jj}|\mathbf{k}\rbrack^{\dot{\be}}_{\ \nn}\epsilon^{\nn\kk}=-\epsilon^{\jj\mm}\lbrack\mathbf{j}|_{\dot{\be}\mm}|\mathbf{k}\rbrack^{\dot{\be}}_{\ \nn}\epsilon^{\nn\kk}= \lbrack\mathbf{j}|_{\dot{\be}}^{\ \jj}|\mathbf{k}\rbrack^{\dot{\be}\kk}.
\end{equation}
Finally, the momentum-insertions of EQ.~\ref{EQSA: SA Mom-Ins} likewise have their counterparts in the standard CSM,
\begin{equation}\label{EQSA: Mom-Ins Equiv.}
    \sm_{xz}\lambda^\dagger_jp_l^-\lambda_k\quad\leftrightarrow\quad \lbrack\mathbf{j}|_{\dot{\be}}^{\ \jj}p_l^{\dot{\be}\al}|\mathbf{k}\rangle_\al^{\ \kk}
\end{equation}
and
\begin{equation}\label{EQSA: Mom-Ins Equiv. 2}
    \sm_{zx}\widetilde{\lambda}_jp_l\lambda_k^-\quad\leftrightarrow\quad \langle\mathbf{j}|^{\al\jj}p_{l,\al\dot{\be}}|\mathbf{k}\rbrack^{\dot{\be}\kk}.
\end{equation}
The power of the APS, and moreso of the Geometric Algebra approach to physics, is manifest in this subsection: The methods of indices and matrix representations can be replaced with geometric, representation-free methods. This dictionary thereby serves as a first step toward a complete understanding of the CSM through Geometric Algebra.

\subsection{Example Calculations}

To demonstrate the SA in squaring constructive scattering amplitudes, three examples will be given. First, the $hf\overline{f}$ (Higgs and two massive fermions) scattering amplitude will be squared, then the $Wq\overline{q}$ (W-boson and two quarks) scattering amplitude, followed by the square of the T-channel photon contribution to the $f_1\overline{f}_1\overline{f}_2f_2$ scattering amplitude. Lastly, the S-channel Higgs-photon crossterm for $f_1\overline{f}_1\overline{f}_2f_2$ will be calculated. As aforementioned, the particle-indices of the SA are redundant, so the first two example calculations will be shown \textit{with and without} indices, while the latter two will only be performed \textit{without} indices.

\subsubsection{$hf\overline{f}$ with Indices}
The coordinate-free form (using $\hat{B}$ rather than $\sm_{xz}$) of the established dictionary is used for a direct translation of the $hf\overline{f}$ scattering amplitude from \cite{Christensen2020}, giving
\begin{equation}\label{EQSA: h -> ff}
    \mathcal{M}_{hf\overline{f}}=-\frac{m_f}{\mathcal{V}}\left(\left[\hat{B}\widetilde{\lambda}_2\lambda_1\right]^{\kk\jj}+\left[\hat{B}^\dagger\lambda_2^\dagger\lambda_1^-\right]^{\kk\jj}\right),
\end{equation}
where $m_f$ is the fermions' mass and $\mathcal{V}$ is the Higgs VEV. Both $\lambda_1$ and $\lambda_2$ are for massive fermions. It is the convention of this paper to write amplitudes like $\mathcal{M}_{hf\overline{f}}$ without indices. This shortens formulas in the CSM. Its Hermitian conjugate is
\begin{equation}\label{EQSA: h -> ff HC}
    \mathcal{M}^\dagger_{hf\overline{f}}=-\frac{m_f}{\mathcal{V}}\left(\left[\lambda_1^\dagger\lambda^-_2\hat{B}^\dagger\right]_{\jj\kk}+\left[\widetilde{\lambda}_1\lambda_2\hat{B}\right]_{\jj\kk}\right).
\end{equation}
The square and sum (using the Einstein convention) over indices has four terms,
\[\begin{aligned}
    \sum_{\jj,\kk}|\mathcal{M}_{h f\overline{f}}|^2&=\frac{m_f^2}{\mathcal{V}^2}\left(\left[\hat{B}\widetilde{\lambda}_2\lambda_1\right]^{\kk\jj}\left[\lambda_1^\dagger\lambda^-_2\hat{B}^\dagger\right]_{\jj\kk}+\left[\hat{B}\widetilde{\lambda}_2\lambda_1\right]^{\kk\jj}\left[\widetilde{\lambda}_1\lambda_2\hat{B}\right]_{\jj\kk}\right) \\
    &+\frac{m_f^2}{\mathcal{V}^2}\left(\left[\hat{B}^\dagger\lambda_2^\dagger\lambda_1^-\right]^{\kk\jj}\left[\lambda_1^\dagger\lambda^-_2\hat{B}^\dagger\right]_{\jj\kk}+\left[\hat{B}^\dagger\lambda_2^\dagger\lambda_1^-\right]^{\kk\jj}\left[\widetilde{\lambda}_1\lambda_2\hat{B}\right]_{\jj\kk}\right),
\end{aligned}\]
which further simplify due to the scalar projection:
\[\begin{aligned}
    &= \frac{m_f^2}{\mathcal{V}^2}\left(2\left<\hat{B}\widetilde{\lambda}_2\lambda_1\lambda_1^\dagger\lambda^-_2\hat{B}^\dagger\right>_{0\oplus 3}+2\left<\hat{B}\widetilde{\lambda}_2\lambda_1\widetilde{\lambda}_1\lambda_2\hat{B}\right>_{0\oplus 3}\right) \\
    &+ \frac{m_f^2}{\mathcal{V}^2}\left(2\left<\hat{B}^\dagger\lambda_2^\dagger\lambda_1^-\lambda_1^\dagger\lambda_2^-\hat{B}^\dagger\right>_{0\oplus 3}+2\left<\hat{B}^\dagger\lambda_2^\dagger\lambda_1^-\widetilde{\lambda}_1\lambda_2\hat{B}\right>_{0\oplus 3}\right) \\
    &= 2\frac{m_f^2}{\mathcal{V}^2}\left(2p_1\sip p_2-2m_f^2\right) \\
    &= 2\frac{m_f^2}{\mathcal{V}^2}\left(m_h^2-4m_f^2\right).
\end{aligned}\]
 The last line was achieved by applying conservation of momentum. This result is in agreement with the standard CSM calculation in \cite{Christensen2020}, as is expected.

\subsubsection{$hf\overline{f}$ without Indices}

The index-free version of EQ.~\ref{EQSA: h -> ff} is
\begin{equation}\label{EQSA: h -> ff 2}
    \mathcal{M}_{hf\overline{f}}=-\frac{m_f}{\mathcal{V}}\left(\left[\hat{B}\widetilde{\lambda}_2\lambda_1\right]+\left[\hat{B}^\dagger\lambda_2^\dagger\lambda_1^-\right]\right),
\end{equation}
and its Hermitian conjugate is
\begin{equation}\label{EQSA: h -> ff HC 2}
    \mathcal{M}^\dagger_{h f\overline{f}}=-\frac{m_f}{\mathcal{V}}\left(\left[\lambda_1^\dagger\lambda^-_2\hat{B}^\dagger\right]+\left[\widetilde{\lambda}_1\lambda_2\hat{B}\right]\right).
\end{equation}
The (summed) sqaure is obtained directly by multiplication of the amplitude with its conjugate:
\[\begin{aligned}
    \sum|\mathcal{M}_{h f\overline{f}}|^2 &=\frac{m_f^2}{\mathcal{V}^2}\left(\left[\hat{B}\widetilde{\lambda}_2\lambda_1\right]\left[\lambda_1^\dagger\lambda^-_2\hat{B}^\dagger\right]+\left[\hat{B}\widetilde{\lambda}_2\lambda_1\right]\left[\widetilde{\lambda}_1\lambda_2\hat{B}\right]\right) \\
    &+\frac{m_f^2}{\mathcal{V}^2}\left(\left[\hat{B}^\dagger\lambda_2^\dagger\lambda_1^-\right]\left[\lambda_1^\dagger\lambda^-_2\hat{B}^\dagger\right]+\left[\hat{B}^\dagger\lambda_2^\dagger\lambda_1^-\right]\left[\widetilde{\lambda}_1\lambda_2\hat{B}\right]\right) \\
    &= \frac{m_f^2}{\mathcal{V}^2}\left(2\left<\hat{B}\widetilde{\lambda}_2\lambda_1\lambda_1^\dagger\lambda^-_2\hat{B}^\dagger\right>_{0\oplus 3}+2\left<\hat{B}\widetilde{\lambda}_2\lambda_1\widetilde{\lambda}_1\lambda_2\hat{B}\right>_{0\oplus 3}\right) \\
    &+ \frac{m_f^2}{\mathcal{V}^2}\left(2\left<\hat{B}^\dagger\lambda_2^\dagger\lambda_1^-\lambda_1^\dagger\lambda_2^-\hat{B}^\dagger\right>_{0\oplus 3}+2\left<\hat{B}^\dagger\lambda_2^\dagger\lambda_1^-\widetilde{\lambda}_1\lambda_2\hat{B}\right>_{0\oplus 3}\right) \\
    &= 2\frac{m_f^2}{\mathcal{V}^2}\left(m_h^2-4m_f^2\right).
\end{aligned}\]
The results are identical.

\subsubsection{$W q\overline{q}$ with Indices}

The unsymmetrized coordinate-free $W q\overline{q}$ scattering amplitude translated from \cite{Christensen2020} is 
\begin{equation}\label{EQSA: W -> qq}
    \mathcal{M}_{W q\overline{q}\ \text{unsym}}=\frac{g_W}{m_W}\left[\hat{B}\widetilde{\lambda}_1\lambda_3\right]^{\jj\Le}\left[\hat{B}^\dagger\lambda_2^\dagger\lambda_3^-\right]^{\kk\mm},
\end{equation}
where $g_W$ is a coupling constant, and $m_W$ is the mass of the W-boson. Both $\lambda_1$ and $\lambda_2$ are quarks, with $\lambda_3$ being the W-boson. The W-boson is a spin-1 particle, and therefore must be symmetrized in its index. Only one of either the amplitude or its conjugate must be symmetrized. For this example, the symmetrized scattering amplitude is used,
\begin{equation}\label{EQSA: W -> qq S}
    \begin{aligned}
        \mathcal{M}_{W q\overline{q}}&=\frac{1}{2}\frac{g_W}{m_W}\left[\hat{B}\widetilde{\lambda}_1\lambda_3\right]^{\jj\Le}\left[\hat{B}^\dagger\lambda_2^\dagger\lambda_3^-\right]^{\kk\mm} \\
        &+\frac{1}{2}\frac{g_W}{m_W}\left[\hat{B}\widetilde{\lambda}_1\lambda_3\right]^{\jj\mm}\left[\hat{B}^\dagger\lambda_2^\dagger\lambda_3^-\right]^{\kk\Le}.
    \end{aligned}
\end{equation}
Now the amplitude is symmetric in $\Le$ and $\mm$. This amplitude's Hermitian conjugate is
\begin{equation}\label{EQSA: W -> qq HC}
    \mathcal{M}^\dagger_{W q\overline{q}}=\frac{g_W}{m_W}\left[\lambda_3^\dagger\lambda_1^-\hat{B}^\dagger\right]_{\Le\jj}\left[\widetilde{\lambda}_3\lambda_2\hat{B}\right]_{\mm\kk}.
\end{equation}
The following square and sum over indices has two terms,
\[\begin{aligned}
    \sum_{\jj,\kk,\Le,\mm}|\mathcal{M}_{W q\overline{q}}|^2 &= \frac{1}{2}\frac{g_W^2}{m_W^2}\left[\hat{B}\widetilde{\lambda}_1\lambda_3\right]^{\jj\Le}\left[\lambda_3^\dagger\lambda_1^-\hat{B}^\dagger\right]_{\Le\jj}\left[\hat{B}^\dagger\lambda_2^\dagger\lambda_3^-\right]^{\kk\mm}\left[\widetilde{\lambda}_3\lambda_2\hat{B}\right]_{\mm\kk}\\
    &+ \frac{1}{2}\frac{g_W^2}{m_W^2}\left[\hat{B}\widetilde{\lambda}_1\lambda_3\right]^{\jj\mm}\left[\widetilde{\lambda}_3\lambda_2\hat{B}\right]_{\mm\kk}\left[\hat{B}^\dagger\lambda_2^\dagger\lambda_3^-\right]^{\kk\Le}\left[\lambda_3^\dagger\lambda_1^-\hat{B}^\dagger\right]_{\Le\jj},
\end{aligned}\]
which map to, and simplify due to, the scalar projections:
\[\begin{aligned}
    &= 2\frac{g_W^2}{m_W^2}\left<\hat{B}\widetilde{\lambda}_1\lambda_3\lambda_3^\dagger\lambda_1^-\hat{B}^\dagger\right>_{0\oplus 3}\left<\hat{B}^\dagger\lambda_2^\dagger\lambda_3^-\widetilde{\lambda}_3\lambda_2\hat{B}\right>_{0\oplus 3} \\
    &+ \frac{g_W^2}{m_W^2}\left<\hat{B}\widetilde{\lambda}_1\lambda_3\widetilde{\lambda}_3\lambda_2\hat{B}\hat{B}^\dagger\lambda_2^\dagger\lambda_3^-\lambda_3^\dagger\lambda_1\hat{B}^\dagger\right>_{0\oplus 3} \\
    &= \frac{g_W^2}{m_W^2}\left(2p_1\sip p_3\, p_2\sip p_3 +m_W^2p_1\sip p_2\right) \\
    &= g_W^2\left(m_W^2-\frac{1}{2}(m_u^2+m_d^2)-\frac{1}{2m_W^2}(m_u^2+m_d^2)^2\right).
\end{aligned}\]
As with the $h f\overline{f}$ case, the last line was achieved by applying conservation of momentum. This result is likewise in agreement with \cite{Christensen2020}. 

\subsubsection{$W q\overline{q}$ without Indices}

The index-free version of EQ.~\ref{EQSA: W -> qq} is
\begin{equation}\label{EQSA: W -> qq 2}
    \mathcal{M}_{W q\overline{q}\ \text{unsym}}=\frac{g_W}{m_W}\left[\hat{B}\widetilde{\lambda}_1\lambda_3\right]\left[\hat{B}^\dagger\lambda_2^\dagger\lambda_3^-\right].
\end{equation}
Its index-free Hermitian conjugate is
\begin{equation}\label{EQSA: W -> qq HC 2}
    \mathcal{M}^\dagger_{W q\overline{q}}=\frac{g_W}{m_W}\left[\lambda_3^\dagger\lambda_1^-\hat{B}^\dagger\right]\left[\widetilde{\lambda}_3\lambda_2\hat{B}\right].
\end{equation}
The index-free symmetrized scattering amplitude is then
\begin{equation}\label{EQSA: W -> qq S 2}
    \begin{aligned}
        \mathcal{M}_{W q\overline{q}}&=\frac{1}{2}\frac{g_W}{m_W}\left[\hat{B}\widetilde{\lambda}_1\lambda_3\right]\left[\hat{B}^\dagger\lambda_2^\dagger\lambda_3^-\right] \\
        &+\frac{1}{2}\frac{g_W}{m_W}\left[\hat{B}^\dagger\lambda_2^\dagger\lambda_3^-\right]\left[\hat{B}\widetilde{\lambda}_1\lambda_3\right].
    \end{aligned}
\end{equation}
 This symmetrized amplitude appears to be strictly equal to the non-symmetrized amplitude. But there is a difference: The reversed second term of EQ.~\ref{EQSA: W -> qq S 2} creates a new term in the square, one that creates a \textit{longer chain} of Lorentz spinors, like what was seen in the square of the indexed amplitudes in the previous subsection. The reversal of this second term in EQ.~\ref{EQSA: W -> qq S 2}, in order to obtain the longer chain, is mnemonic. This implies that the term could have been another $\left[\hat{B}\widetilde{\lambda}_1\lambda_3\right]\left[\hat{B}^\dagger\lambda_2^\dagger\lambda_3^-\right]/2$, as long as the longer chain were to be constructed anyway. Therefore symmetrization is about introducing a factor of $1/2$ per spin-$1$ particle, and accounting for the new chains formed. Now, calculating the index-free squared amplitude:
\[\begin{aligned}
    \sum|\mathcal{M}_{W q\overline{q}}|^2&=\frac{1}{2}\frac{g_W^2}{m_W^2}\left[\hat{B}\widetilde{\lambda}_1\lambda_3\right]\left[\lambda_3^\dagger\lambda_1^-\hat{B}^\dagger\right]\left[\hat{B}^\dagger\lambda_2^\dagger\lambda_3^-\right]\left[\widetilde{\lambda}_3\lambda_2\hat{B}\right] \\
    &+\frac{1}{2}\frac{g_W^2}{m_W^2}\left[\hat{B}^\dagger\lambda_2^\dagger\lambda_3^-\right]\left[\lambda_3^\dagger\lambda_1^-\hat{B}^\dagger\right]\left[\hat{B}\widetilde{\lambda}_1\lambda_3\right]\left[\widetilde{\lambda}_3\lambda_2\hat{B}\right] \\
    &=\frac{1}{2}\frac{g_W^2}{m_W^2}4\left<\hat{B}\widetilde{\lambda}_1\lambda_3\lambda_3^\dagger\lambda_1^-\hat{B}^\dagger\right>_{0\oplus 3}\left<\hat{B}^\dagger\lambda_2^\dagger\lambda_3^-\widetilde{\lambda}_3\lambda_2\hat{B}\right>_{0\oplus 3} \\
    &+ \frac{1}{2}\frac{g_W^2}{m_W^2}2\left<\hat{B}^\dagger\lambda_2^\dagger\lambda_3^-\lambda_3^\dagger\lambda_1^-\hat{B}^\dagger\hat{B}\widetilde{\lambda}_1\lambda_3\widetilde{\lambda}_3\lambda_2\hat{B}\right>_{0\oplus 3} \\
    &= g_W^2\left(m_W^2-\frac{1}{2}(m_u^2+m_d^2)-\frac{1}{2m_W^2}(m_u^2+m_d^2)^2\right).
\end{aligned}\]
The results are once more identical. But clearly, there is a new and longer chain, $\lambda_2\rightarrow\lambda_3\rightarrow\lambda_1\rightarrow\lambda_3$, that would not have been obtained without symmetrization.

\subsubsection{T-Channel Photon Contribution to $f_1\overline{f}_1\overline{f}_2f_2$}

Translating from \cite{3Christensen2024}, the \textit{T-channel photon contribution} to the $f_1\overline{f}_1\overline{f}_2f_2$ scattering amplitude is 
\begin{equation}\label{EQSA: g T-channel}
    \begin{aligned}
        \mathcal{M}_{\gm T}&=-\kappa_{\gm T}\left[\hat{B}^\dagger\lambda_1^\dagger\lambda_4^-\right]^{\jj\mm}\left[\hat{B}\widetilde{\lambda}_2\lambda_3\right]^{\kk\Le} \\
        &+\kappa_{\gm T}\left[\hat{B}^\dagger\lambda_1^\dagger\lambda_2^-\right]^{\jj\kk}\left[\hat{B}\widetilde{\lambda}_3\lambda_4\right]^{\Le\mm} \\
        &-\kappa_{\gm T}\left[\hat{B}\widetilde{\lambda}_1\lambda_4\right]^{\jj\mm}\left[\hat{B}^\dagger\lambda_2^\dagger\lambda_3^-\right]^{\kk\Le} \\
        &+ \kappa_{\gm T}\left[\hat{B}\widetilde{\lambda}_1\lambda_2\right]^{\jj\kk}\left[\hat{B}^\dagger\lambda_3^\dagger\lambda_4^-\right]^{\Le\mm},
    \end{aligned}
\end{equation}
for $\kappa_{\gm T}=2Q_1Q_2e^2/t$, where $Q_j$ is the charge of $f_j$ in units of $e$, and $t$ is a T-channel Mandelstahm variable. The index-free version is
\begin{equation}\label{EQSA: g T-channel 2}
    \begin{aligned}
        \mathcal{M}_{\gm T}&=-\kappa_{\gm T}\left[\hat{B}^\dagger\lambda_1^\dagger\lambda_4^-\right]\left[\hat{B}\widetilde{\lambda}_2\lambda_3\right] 
        +\kappa_{\gm T}\left[\hat{B}^\dagger\lambda_1^\dagger\lambda_2^-\right]\left[\hat{B}\widetilde{\lambda}_3\lambda_4\right] \\
        &-\kappa_{\gm T}\left[\hat{B}\widetilde{\lambda}_1\lambda_4\right]\left[\hat{B}^\dagger\lambda_2^\dagger\lambda_3^-\right] + \kappa_{\gm T}\left[\hat{B}\widetilde{\lambda}_1\lambda_2\right]\left[\hat{B}^\dagger\lambda_3^\dagger\lambda_4^-\right],
    \end{aligned}
\end{equation}
and the index-free Hermitian conjugate is 
\begin{equation}\label{EQSA: g T-channel HC}
    \begin{aligned}
        \mathcal{M}^\dagger_{\gm T}&=-\kappa_{\gm T}\left[\widetilde{\lambda}_4\lambda_1\hat{B}\right]\left[\lambda^\dagger_3\lambda_2^-\hat{B}^\dagger\right] +\kappa_{\gm T}\left[\widetilde{\lambda}_2\lambda_1\hat{B}\right]\left[\lambda_4^\dagger\lambda_3^-\hat{B}^\dagger\right]\\
        &-\kappa_{\gm T}\left[\lambda_4^\dagger\lambda_1^-\hat{B}^\dagger\right]\left[\widetilde{\lambda}_3\lambda_2\hat{B}\right] + \kappa_{\gm T}\left[\lambda_2^\dagger\lambda_1^-\hat{B}^\dagger\right]\left[\widetilde{\lambda}_4\lambda_3\hat{B}\right].\\
    \end{aligned}
\end{equation}
All particles are spin-$\frac{1}{2}$, with $\lambda_1$ and $\lambda_2$ corresponding to $f_1$ and $\overline{f}_1$, and with $\lambda_3$ and $\lambda_4$ corresponding to $f_2$ and $\overline{f}_2$. The (summed) square of the amplitude is once more obtained via direct multiplication of the amplitude and its conjugate. The first four terms of the squared amplitude are
\[\begin{aligned}
    &\kappa_{\gm T}^2\left[\hat{B}^\dagger\lambda_1^\dagger\lambda_4^-\right]\left[\widetilde{\lambda}_4\lambda_1\hat{B}\right]\left[\hat{B}\widetilde{\lambda}_2\lambda_3\right] \left[\lambda^\dagger_3\lambda_2^-\hat{B}^\dagger\right] \\
        &-\kappa_{\gm T}^2\left[\hat{B}^\dagger\lambda_1^\dagger\lambda_4^-\right]\left[\widetilde{\lambda}_2\lambda_1\hat{B}\right]\left[\hat{B}\widetilde{\lambda}_2\lambda_3\right]\left[\lambda_4^\dagger\lambda_3^-\hat{B}^\dagger\right] \\
        &+ \kappa_{\gm T}^2\left[\hat{B}^\dagger\lambda_1^\dagger\lambda_4^-\right] \left[\lambda_4^\dagger\lambda_1^-\hat{B}^\dagger\right]\left[\hat{B}\widetilde{\lambda}_2\lambda_3\right]\left[\widetilde{\lambda}_3\lambda_2\hat{B}\right]\\
        &-\kappa_{\gm T}^2 \left[\hat{B}^\dagger\lambda_1^\dagger\lambda_4^-\right]\left[\lambda_2^\dagger\lambda_1^-\hat{B}^\dagger\right]
        \left[\hat{B}\widetilde{\lambda}_2\lambda_3\right]\left[\widetilde{\lambda}_4\lambda_3\hat{B}\right] .
\end{aligned}\]
Notice that the second and fourth lines do not have properly ordered particle-labels. To remedy this, the antisymmetry of particle exchange in EQ.~\ref{EQSA: Antisymmetric Particle Exchange} is applied. This results in the four terms simplifying to scalar projections:
\[\begin{aligned}
    &= 4\kappa_{\gm T}^2\left<\hat{B}^\dagger\lambda_1^\dagger\lambda_4^-\widetilde{\lambda}_4\lambda_1\hat{B}\right>_{0\oplus 3}\left<\hat{B}\widetilde{\lambda}_2\lambda_3\lambda_3^\dagger\lambda_2^-\hat{B}^\dagger\right>_{0\oplus 3} \\
    &+2\kappa_{\gm T}^2\left<\hat{B}^\dagger\lambda_4^\dagger\lambda_1^-\widetilde{\lambda}_1\lambda_2\hat{B}^2\widetilde{\lambda}_2\lambda_3\lambda_3^\dagger\lambda_4^-\hat{B}^\dagger\right>_{0\oplus 3} \\
    &+4\kappa_{\gm T}^2\left<\hat{B}^\dagger\lambda_1^\dagger\lambda_4^-\lambda_4^\dagger\lambda_1^-\hat{B}^\dagger\right>_{0\oplus 3}\left<\hat{B}\widetilde{\lambda}_2\lambda_3\widetilde{\lambda}_3\lambda_2\hat{B}\right>_{0\oplus 3} \\
    &+ 2\kappa_{\gm T}^2\left<\hat{B}^\dagger\lambda_4^\dagger\lambda_1^-\lambda_1^\dagger\lambda_2^-\hat{B}^\dagger\hat{B}\widetilde{\lambda}_2\lambda_3\widetilde{\lambda}_3\lambda_4\hat{B}\right>_{0\oplus 3} \\
    &= 2\kappa_{\gm T}^2(2p_1\sip p_4\,p_2\sip p_3+p_1\sip p_3m_2m_4+3m_1m_2m_3m_4+p_2\sip p_4m_1m_3).
\end{aligned}\]
The remaining twelve terms in the squared amplitude are obtained identically, and the final result is
\[\begin{aligned}
    \sum|\mathcal{M}_{(\gm/g)T}|^2&= 8\kappa_{\gm T}^2(p_1\sip p_4\, p_2\sip p_3+p_1\sip p_2\, p_3\sip p_4) \\ 
    &+ 8\kappa_{\gm T}^2(p_1\sip p_3m_2m_4+p_2\sip p_4m_1m_3+2m_1m_2m_3m_4).
\end{aligned}\]
This has been further simplified and verified by the authors. 

\subsubsection{S-Channel Higgs-Photon Crossterm for $f_1\overline{f}_1\overline{f}_2f_2$}
Translating from \cite{3Christensen2024}, the \textit{S-channel Higgs contribution} to the $f_1\overline{f}_1\overline{f}_2f_2$ scattering amplitude is 
\begin{equation}\label{EQSA: S-channel higgs}
    \begin{aligned}
        \mathcal{M}_{hS}=\kappa_{hS}&\left(\left[\hat{B}\widetilde{\lambda}_1\lambda_2\right]^{\jj\kk}+\left[\hat{B}^\dagger\lambda_1^\dagger\lambda_2^-\right]^{\jj\kk}\right) \\
        &\left(\left[\hat{B}\widetilde{\lambda}_3\lambda_4\right]^{\Le\mm}+\left[\hat{B}^\dagger\lambda_3^\dagger\lambda_4^-\right]^{\Le\mm}\right)
    \end{aligned}
\end{equation}
for the coefficient  $\kappa_{hS}=e^2m_1m_2/4m_W^2s_W^2(s-m^2_h)$. Here $e$ is the unit of charge, $m_j$ is the mass of $f_j$, $m_W$ is the mass of the W-boson, $m_h$ is the mass of the Higgs boson, $s_W=\sin{\theta_W}$ is the sine of the Weinberg angle, and $s$ is an S-channel Mandelstahm variable. The \textit{S-channel photon contribution} is likewise
\begin{equation}\label{EQSA: S-channel g}
    \begin{aligned}
        \mathcal{M}_{\gm S}&=\kappa_{\gm S}\left[\hat{B}^\dagger\lambda_1^\dagger\lambda_4^-\right]^{\jj\mm}\left[\hat{B}\widetilde{\lambda}_2\lambda_3\right]^{\kk\Le} \\
        &+\kappa_{\gm S}\left[\hat{B}^\dagger\lambda_1^\dagger\lambda_3^-\right]^{\jj\Le}\left[\hat{B}\widetilde{\lambda}_2\lambda_4\right]^{\kk\mm} \\
        &+\kappa_{\gm S}\left[\hat{B}\widetilde{\lambda}_1\lambda_4\right]^{\jj\mm}\left[\hat{B}^\dagger\lambda_2^\dagger\lambda_3^-\right]^{\kk\Le} \\
        &+ \kappa_{\gm S}\left[\hat{B}\widetilde{\lambda}_1\lambda_3\right]^{\jj\Le}\left[\hat{B}^\dagger\lambda_2^\dagger\lambda_4^-\right]^{\kk\mm},
    \end{aligned}
\end{equation}
for $\kappa_{\gm S}=-2Q_1Q_2e^2/s$. The index-free amplitudes for both the Higgs and photon contributions are respectively
\begin{equation}\label{EQSA: S-channel higgs 2}
        \mathcal{M}_{hS}=\kappa_{hS}\left(\left[\hat{B}\widetilde{\lambda}_1\lambda_2\right]+\left[\hat{B}^\dagger\lambda_1^\dagger\lambda_2^-\right]\right)
        \left(\left[\hat{B}\widetilde{\lambda}_3\lambda_4\right]+\left[\hat{B}^\dagger\lambda_3^\dagger\lambda_4^-\right]\right)
\end{equation}
and
\begin{equation}\label{EQSA: T-channel g 2}
    \begin{aligned}
        \mathcal{M}_{\gm S}&=\kappa_{\gm S}\left[\hat{B}^\dagger\lambda_1^\dagger\lambda_4^-\right]\left[\hat{B}\widetilde{\lambda}_2\lambda_3\right]
        +\kappa_{\gm S}\left[\hat{B}^\dagger\lambda_1^\dagger\lambda_3^-\right]\left[\hat{B}\widetilde{\lambda}_2\lambda_4\right] \\
        &+\kappa_{\gm S}\left[\hat{B}\widetilde{\lambda}_1\lambda_4\right]\left[\hat{B}^\dagger\lambda_2^\dagger\lambda_3^-\right] + \kappa_{\gm S}\left[\hat{B}\widetilde{\lambda}_1\lambda_3\right]\left[\hat{B}^\dagger\lambda_2^\dagger\lambda_4^-\right],
    \end{aligned}
\end{equation}
with 
\begin{equation}\label{EQSA: S-channel higgs HC}
        \mathcal{M}_{hS}^\dagger=\kappa_{hS}\left(\left[\lambda_2^\dagger\lambda_1^-\hat{B}^\dagger\right]+\left[\widetilde{\lambda}_2\lambda_1\hat{B}\right]\right)
        \left(\left[\lambda_4^\dagger\lambda_3^-\hat{B}^\dagger\right]+\left[\widetilde{\lambda}_4\lambda_3\hat{B}\right]\right)
\end{equation}
being the Hermitian conjugate of EQ.~\ref{EQSA: S-channel higgs 2}. There are sixteen terms in total within the crossterm $\sum\mathcal{M}_{\gm S}\mathcal{M}_{hS}^\dagger$. The first term is
\[
\kappa_{\gm S}\kappa_{hS}\left[\hat{B}^\dagger\lambda_1^\dagger\lambda_4^-\right]\left[\lambda_4^\dagger\lambda_3^-\hat{B}^\dagger\right]\left[\hat{B}\widetilde{\lambda}_2\lambda_3\right]\left[\lambda_2^\dagger\lambda_1^-\hat{B}^\dagger\right].
\]
The particle-labels are once more not properly ordered, so the antisymmetry in EQ.~\ref{EQSA: Antisymmetric Particle Exchange} is again used to obtain
\[
\begin{aligned}
    \kappa_{\gm S}\kappa_{hS}&\left[\hat{B}^\dagger\lambda_1^\dagger\lambda_4^-\right]\left[\lambda_4^\dagger\lambda_3^-\hat{B}^\dagger\right]\left[\hat{B}^\dagger\widetilde{\lambda}_3\lambda_2\right]\left[\lambda_2^\dagger\lambda_1^-\hat{B}^\dagger\right] \\
    &= 2\kappa_{\gm S}\kappa_{hS}\left<\hat{B}^\dagger\lambda_1^\dagger\lambda_4^-\lambda_4^\dagger\lambda_3^-\hat{B}^\dagger\hat{B}^\dagger\widetilde{\lambda}_3\lambda_2\lambda_2^\dagger\lambda_1^-\hat{B}^\dagger\right>_{0\oplus 3} \\
    &= 2\kappa_{\gm S}\kappa_{hS}m_1m_4p_2\sip p_3.
\end{aligned}
\]
The remaining fifteen terms are found in the same manner, with
\[
\sum\mathcal{M}_{\gm S}\mathcal{M}_{hS}^\dagger = 8\kappa_{\gm S}\kappa_{hS}(m_1m_4p_2\sip p_3-m_2m_4p_1\sip p_3-m_1m_3p_2\sip p_4+m_2m_3p_1\sip p_4)
\]
as the resulting crossterm. This has been further simplified and verified by the authors.

\section{Discussion}

Compared to the traditional methods of the Constructive Standard Model (CSM), the Geometric Algebra approach within the Algebra of Physical Space (APS) has manifest power. The structure and methods involving indices and matrices can be analyzed with geometric, representation-free methods. As an unintended side-effect of such an analysis, the geometry of the APS has given two new insights into the CSM. The first is the fact that the \textit{spin spinors} of the CSM, discussed in SEC.~\ref{SA}, are simply \textit{Lorentz spinors} (scaled Lorentz rotors) and therefore cannot hold any particle-spin information. Included in this first insight is the identification of a new type of spin spinor not discussed previously within the traditional CSM, but well-known within the APS: The Hermitian $\lambda_p$ of EQ.~\ref{EQSF: p = LpLp}, in this paper called an \textit{achiral} Lorentz spinor. The second insight is the discovery that the CSM adheres to the \textit{ray spinor structure}, presented in SEC.~\ref{SF}. Algebraically, this connects the CSM to other formalisms like the \textit{Vaz-da Rocha formalism} in \cite{Vaz2019}, to the \textit{Penrose-Rindler formalism} in \cite{Penrose1986}, and to the \textit{two-spinor calculus} in \cite{Doran2003}. Geometrically, this insight demonstrates a way to view the CSM through the geometry of forward-oriented and backward-oriented lightrays on the lightcone. 

On top of these insights, the results of the CSM were replicated in the massive cases. This was made possible by the introduction of the \textit{Scattering Algebra} (SA) in SEC.~\ref{SA}, which is a novel tool within the APS that establishes isomorphism between the traditional CSM and the APS approach. However, this paper only explored massive cases. A line of further research is the replication of the CSM's massless cases using the APS and the SA. Closely related to this is the exploration of Wigner little group methods inside the APS, and the development of a method for amplitude-construction within the APS and SA.


\section*{Acknowledgements}
The authors would like to thank Edward Corbett, Martin Roelfs, and Hamish Todd for helpful discussions. The authors would also like to express their gratitude to the American people for partially supporting this work through the National Science Foundation under Grant No. PHY-2411482.

\subsubsection*{Data Availability Statement}
This paper contains no results found through data analysis or any like method.

\section*{Original Research}
This paper is a pre-final-acceptance-proof version of a publication in \textbf{Springer Link} \textit{Advances in Applied Clifford Algebras}. The official DOI is \textit{10.1007/s00006-025-01435-1}. Please cite the official release.


\appendix

\section{Chirality of Lorentz Spinors}\label{A0.5}

Consider a Lorentz transformation which boosts in the $\hat{c}\in\APS^1$ direction and rotates about the axis, $i\hat{c}\in\APS^2$, parallel to the boost:
\begin{equation}\label{EQLC: Special RL}
    \Lambda_c=e^{\frac{1}{2}(\zeta_1-i\zeta_2)\hat{c}}.
\end{equation}
Here $\zeta_j\in\mathrm{Center}[\APS]=\APS^{0\oplus3}\approx\mathbb{C}$ is allowed by the fact that $\hat{c}$ and $i\hat{c}$ are dual to one another (through multiplication by $\pm i$). This is a completely general way of parameterizing a Lorentz transformation which rotates about the boost direction. 

A \textit{left-chiral} transformation involves $\Lambda_c$, and choosing $\zeta_2=i\zeta_1$. Under a left-chiral transformation, it is seen that of the Lorentz spinors $\lambda$, $\widetilde{\lambda}$, $\lambda^\dagger$, and $\lambda^-$, which are determined by $\lambda=\sqrt{mc}R_pL_a$ in EQ.~\ref{EQSF: p = RLLR}, only the first two spinors transform whereas the latter two remain the same:
\begin{equation}\label{EQLC: Left}
    \lambda\mapsto e^{\zeta_1\hat{c}}\lambda,\quad\widetilde{\lambda}\mapsto\widetilde{\lambda}e^{-\zeta_1\hat{c}},\quad\lambda^\dagger\mapsto\lambda^\dagger,\quad\lambda^-\mapsto\lambda^-.
\end{equation}
Likewise, a \textit{right-chiral} transformation involves $\Lambda_c$ with $\zeta_2=-i\zeta_1$. Under a right-chiral transformation, the former two spinors stay the same while the latter two transform:
\begin{equation}\label{EQLC: Right}
    \lambda\mapsto \lambda,\quad\widetilde{\lambda}\mapsto\widetilde{\lambda},\quad\lambda^\dagger\mapsto\lambda^\dagger e^{\zeta_1\hat{c}},\quad\lambda^-\mapsto e^{-\zeta_1\hat{c}}\lambda^- .
\end{equation}
Therefore $\lambda$ and $\widetilde{\lambda}$ are said to transform \textit{left-chirally}, while $\lambda^\dagger$ and $\lambda^-$ are said to transform \textit{right-chirally}. These transformations are necessary but not sufficient to declare the spinors' chiralities, however. For example, consider the Lorentz spinor $\lambda_p=\sqrt{mc}L_p$ of EQ.~\ref{EQSF: p = LpLp}. This spinor and its conjugates obey EQ.~\ref{EQLC: Left} and EQ.~\ref{EQLC: Right}. This spinor is also Hermitian, meaning $\lambda_p=\lambda_p^\dagger$. A Lorentz spinor which transforms left-chirally is equal to a Lorentz spinor which transforms right-chirally. In this case, it is implied that $\lambda_p$ is \textit{achiral}. Thus the additional condition for sufficiency when determining a Lorentz spinor's chirality is that, without loss of generality, $\lambda\neq\lambda^\dagger$. In short, $\lambda$ and $\lambda^\dagger$ are respectively left-chiral and right-chiral \textit{if and only if} $\lambda$ transforms left-chirally while $\lambda^\dagger$ transforms right-chirally, and $\lambda\neq\lambda^\dagger$. An identical statement can be made for $\widetilde{\lambda}$ and $\lambda^-$. It is for this reason that $\lambda=\sqrt{mc}R_pL_a$ (and its conjugates) are called \textit{chiral} Lorentz spinors. 

It should be noted that the Lorentz product of EQ.~\ref{EQSA: LI Product} is called left-chiral because the product remains invariant non-trivially under left-chiral transformations. Likewise, the Lorentz product of EQ.~\ref{EQSA: RC LI Product} is called right-chiral because the product remains invariant non-trivially under right-chiral transformations. Under right-chiral transformations, the left-chiral Lorentz product remains \textit{trivially} invariant. The same is true for the right-chiral Lorentz product under left-chiral transformations. It is for these reasons that a distinction between left-chirality and right-chirality can be made for the Lorentz products.

It was noted in SEC.~\ref{SA} that the left-chiral and right-chiral Lorentz products are related to one another via the (light)ray spinor structure of SEC.~\ref{SF}. Specifically, using $\lambda_k\mapsto\lambda_k^-$ and $\hat{B}\widetilde{\lambda}_j\mapsto-(\hat{B}\widetilde{\lambda}_j)^-$. This demonstrates that the difference between the chiral Lorentz products is held by the ray spinor structure, independent of the chirality of the spinors themselves.

\section{Table of Concurrence}\label{A0.75}

\bigskip
\noindent
\begin{center}
\begin{tabular}{llll}
\toprule
\textbf{Object} & \textbf{Traditional} & \textbf{Relation} & \textbf{SA/APS} \\
\midrule
Right-Angle Spinor & $|\mathbf{j}\rangle_\al^{\ \jj}$ & $\equiv$ & $\la=\sqrt{mc}R_pL_a$\\
Left-Angle Spinor (Transpose) & $\left(\langle\mathbf{j}|^{\al\jj}\right)^T$ & $\equiv$ & $\hat{B}\widetilde{\la}$\\
Right-Square Spinor & $|\mathbf{j}\rbrack^{\dot{\be}}_{\ \jj}$ & $\equiv$ & $\la^-\hat{B}^\dagg$\\
Left-Square Spinor (Transpose) & $\left(\lbrack\mathbf{j}|_{\dot{\be}\jj}\right)^T$ & $\equiv$ & $\la^\dagg$\\
(Raising) Epsilon Tensor & $\epsilon^{\jj\kk}$ & $=$ & $\upSAb{\hat{B}^\dagg}{\jj}{\kk}$\\
(Lowering) Epsilon Tensor & $\epsilon_{\jj\kk}$ & $=$ & $\downSAb{\hat{B}}{\kk}{\jj}$\\
(Left) Lorentz Product & $\langle\mathbf{j}|^{\al\jj}|\mathbf{k}\rangle_\al^{\ \kk}$ & $=$ & $\upSAb{\LP{j}{k}}{\jj}{\kk}$ \\
(Right) Lorentz Product & $\lbrack\mathbf{j}|_{\dot{\be}}^{\ \jj}|\mathbf{k}\rbrack^{\dot{\be}\kk}$ & $=$ & $\upSAb{\mLP{j}{k}}{\jj}{\kk}$\\
(First) Momentum Insertion & $\lbrack\mathbf{j}|_{\dot{\be}}^{\ \jj}p_l^{\dot{\be}\al}|\mathbf{k}\rangle_\al^{\ \kk}$ & $=$ & $\upSAb{\hat{B}\la_j^\dagg p_l^-\la_k}{\jj}{\kk}$\\
(Second) Momentum Insertion & $\langle\mathbf{j}|^{\al\jj}p_{l,\al\dot{\be}}|\mathbf{k}\rbrack^{\dot{\be}\kk}$ & $=$ & $\upSAb{\hat{B}^\dagg\widetilde{\la}_j p_l\la_k^-}{\jj}{\kk}$\\
\bottomrule
\end{tabular}
\end{center}
\bigskip

\section{Translated Constructive Amplitudes}\label{A1}

\subsection{$2$-Body Decay Amplitudes}

There are three massive fermionic $2$-body decay amplitudes in the Standard Model, $hf\overline{f}$ (Higgs and two massive fermions), $Wq\overline{q}$ (W-boson and two quarks), and $Z f\overline{f}$ (Z-boson and two massive fermions). 

\subsubsection{$h f\overline{f}$}
The constructive scattering amplitude for a \textit{Higgs and two massive fermions} is given by EQ.~\ref{EQSA: h -> ff}.

\subsubsection{$W q\overline{q}$}
The constructive scattering amplitude for a \textit{W-boson and two quarks} is given by EQ.~\ref{EQSA: W -> qq}. 

\subsubsection{$Z f\overline{f}$}
Translating from \cite{Christensen2020}, the constructive scattering amplitude for a \textit{Z-boson and two massive fermions} is 
\begin{equation}\label{EQCA: Z -> ff}
    \begin{aligned}
        \mathcal{M}_{Z f\overline{f}} &= \frac{g_L}{m_Z}\left[\hat{B}\widetilde{\lambda}_1\lambda_3\right]^{\jj\kk}\left[\hat{B}^\dagger\lambda_2^\dagger\lambda_3^-\right]^{\Le\mm} \\
        &+ \frac{g_R}{m_Z}\left[\hat{B}^\dagger\lambda_1^\dagger\lambda_3^-\right]^{\jj\kk}\left[\hat{B}\widetilde{\lambda}_2\lambda_3\right]^{\Le\mm},
    \end{aligned}
\end{equation}
where the $g_{L/R}$ are left-chirality and right-chirality coupling constants, and $m_Z$ is the mass of the Z-boson. Both $\lambda_1$ and $\lambda_2$ are massive fermions, and $\lambda_3$ is the Z-boson. Since the Z-boson is spin-1, the symmetrization techniques shown in SEC.~\ref{SA} for the $W q\overline{q}$ amplitude must be employed when squaring. The $Z f\overline{f}$ amplitude has been confirmed to square to the value calculated in both the Constructive Standard Model (CSM) and the SM.

\subsection{$f_1\overline{f}_1 \overline{f}_2f_2$ Scattering Amplitudes}

For \textit{two pairs of massive fermions} there are contributions to the scattering amplitude from photons, Higgs-bosons, Z-bosons, and W-bosons. These come in S-channels and T-channels for a total of seven terms. When squaring the full amplitude, 
\begin{equation}\label{EQCA: Full ff -> ff}
    \begin{aligned}
        \mathcal{M}_{f_1\overline{f}_1\overline{f}_2f_2} = \mathcal{M}_{\gm S}+\mathcal{M}_{\gm T}+\mathcal{M}_{hS}+\mathcal{M}_{hT} \\
        +\mathcal{M}_{ZS}+\mathcal{M}_{ZT}+\mathcal{M}_{WT}
    \end{aligned}
\end{equation}
there are squared terms and \textit{crossterms}. Both squared terms and crossterms have been confirmed to agree with the values calculated in both the CSM and the FSM.

For the upcoming constructive amplitude contributions, the Lorentz spinors $\lambda_1$ and $\lambda_2$ correspond to $f_1$ and $\overline{f}_1$, while the Lorentz spinors $\lambda_3$ and $\lambda_4$ correspond to $f_2$ and $\overline{f}_2$. 

\subsubsection{S-Channel Photon Contribution}

The constructive scattering amplitude for the \textit{S-channel photon contribution} is given by EQ.~\ref{EQSA: S-channel g}. 

\subsubsection{T-Channel Photon Contribution}

The constructive scattering amplitude for the \textit{T-channel photon contribution} is given by EQ.~\ref{EQSA: g T-channel}.

\subsubsection{S-Channel Higgs Contribution}

The constructive scattering amplitude for the \textit{S-channel Higgs contribution} is given by EQ.~\ref{EQSA: S-channel higgs}.

\subsubsection{T-Channel Higgs Contribution}

The T-channel Higgs contribution only appears when $f_2=f_1$. Translating from \cite{3Christensen2024}, the \textit{T-channel Higgs contribution} is
\begin{equation}\label{EQCA: h T-channel}
    \begin{aligned}
        \mathcal{M}_{hT}&=\kappa_{hT}\left(\left[\hat{B}\widetilde{\lambda}_1\lambda_3\right]^{\jj\kk}+\left[\hat{B}^\dagger\lambda_1^\dagger\lambda_3^-\right]^{\jj\kk}\right)\left[\hat{B}\widetilde{\lambda}_2\lambda_4\right]^{\Le\mm} \\
        &+ \kappa_{hT}\left(\left[\hat{B}\widetilde{\lambda}_1\lambda_3\right]^{\jj\kk}+\left[\hat{B}^\dagger\lambda_1^\dagger\lambda_3^-\right]^{\jj\kk}\right)\left[\hat{B}^\dagger\lambda_2^\dagger\lambda_4^-\right]^{\Le\mm},
    \end{aligned}
\end{equation}
where $\kappa_{hT}=-\kappa_{hS}(s-m_h^2)/(t-m_h^2)$, with $t$ being a T-channel Mandelstahm variable.

\subsubsection{S-Channel Z-Boson Contribution}

From \cite{3Christensen2024}, the \textit{S-channel Z-boson contribution} is
\begin{equation}\label{EQCA: Z S-channel}
    \begin{aligned}
        \mathcal{M}_{ZS}&=\kappa_{ZS1}\left(\left[\hat{B}\widetilde{\lambda}_1\lambda_2\right]^{\jj\kk}-\left[\hat{B}^\dagger\lambda_1^\dagger\lambda_2\right]^{\jj\kk}\right)\left[\hat{B}\widetilde{\lambda}_3\lambda_4\right]^{\Le\mm} \\
        &-\kappa_{ZS1}\left(\left[\hat{B}\widetilde{\lambda}_1\lambda_2\right]^{\jj\kk}-\left[\hat{B}^\dagger\lambda_1^\dagger\lambda_2^-\right]^{\jj\kk}\right)\left[\hat{B}^\dagger\lambda_3^\dagger\lambda_4^-\right]^{\Le\mm} \\
        &+\kappa_{ZS2}g_{R1}g_{R2}\left[\hat{{B}^\dagger}\lambda_1^\dagger\lambda_4^-\right]^{\jj\mm}\left[\hat{B}\widetilde{\lambda}_2\lambda_3\right]^{\kk\Le} \\
        &+\kappa_{ZS2}g_{L2}g_{R1}\left[\hat{{B}^\dagger}\lambda_1^\dagger\lambda_3^-\right]^{\jj\Le}\left[\hat{B}\widetilde{\lambda}_2\lambda_4\right]^{\kk\mm} \\
        &+\kappa_{ZS2}g_{L1}g_{R2}\left[\hat{B}\widetilde{\lambda}_1\lambda_3\right]^{\jj\Le}\left[\hat{B}^\dagger\lambda_2^\dagger\lambda_4^-\right]^{\kk\mm} \\
        &+\kappa_{ZS2}g_{L1}g_{L2}\left[\hat{B}\widetilde{\lambda}_1\lambda_4\right]^{\jj\mm}\left[\hat{B}^\dagger\lambda_2^\dagger\lambda_3^-\right]^{\kk\Le},
    \end{aligned}
\end{equation}
where $\kappa_{ZS1}=-e^2m_1m_2(g_{L1}-g_{R1})(g_{L2}-g_{R2})/4m_W^2s_W^2(s-m_Z^2)$ and $\kappa_{ZS2}=-e^2/2c_W^2s_W^2(s-m_Z^2)$. Here $g_{(L/R)j}$ are left and right-chirality coupling constants for $f_j$, $m_Z$ is the mass of the Z-boson, and $c_W=\cos{\theta_W}$.

\subsubsection{T-Channel Z-Boson Contribution} 

The \textit{T-channel Z-boson contribution} looks almost identical to EQ.~\ref{EQCA: Z S-channel}. From \cite{3Christensen2024},
\begin{equation}\label{EQCA: Z T-channel}
    \begin{aligned}
        \mathcal{M}_{ZT}&=\kappa_{ZT1}\left(\left[\hat{B}\widetilde{\lambda}_1\lambda_3\right]^{\jj\Le}-\left[\hat{B}^\dagger\lambda_1^\dagger\lambda_3\right]^{\jj\Le}\right)\left[\hat{B}\widetilde{\lambda}_2\lambda_4\right]^{\kk\mm} \\
        &-\kappa_{ZT1}\left(\left[\hat{B}\widetilde{\lambda}_1\lambda_3\right]^{\jj\Le}-\left[\hat{B}^\dagger\lambda_1^\dagger\lambda_3^-\right]^{\jj\Le}\right)\left[\hat{B}^\dagger\lambda_2^\dagger\lambda_4^-\right]^{\kk\mm} \\
        &-\kappa_{ZT2}g_{R1}^2\left[\hat{{B}^\dagger}\lambda_1^\dagger\lambda_4^-\right]^{\jj\mm}\left[\hat{B}\widetilde{\lambda}_2\lambda_3\right]^{\kk\Le} \\
        &+\kappa_{ZT2}g_{L1}g_{R1}\left[\hat{{B}^\dagger}\lambda_1^\dagger\lambda_2^-\right]^{\jj\kk}\left[\hat{B}\widetilde{\lambda}_3\lambda_4\right]^{\Le\mm} \\
        &+\kappa_{ZT2}g_{L1}g_{R1}\left[\hat{B}\widetilde{\lambda}_1\lambda_2\right]^{\jj\kk}\left[\hat{B}^\dagger\lambda_3^\dagger\lambda_4^-\right]^{\Le\mm} \\
        &-\kappa_{ZT2}g_{L1}^2\left[\hat{B}\widetilde{\lambda}_1\lambda_4\right]^{\jj\mm}\left[\hat{B}^\dagger\lambda_2^\dagger\lambda_3^-\right]^{\kk\Le},
    \end{aligned}
\end{equation}
where $\kappa_{ZT1}=-\kappa_{ZS1}(s-m_Z^2)(g_{L1}-g_{R1})/(t-m_Z^2)(g_{L2}-g_{R2})$ and $\kappa_{ZT2}=-\kappa_{ZS2}(s-m_Z^2)/(t-m_Z^2)$.

\subsubsection{T-Channel W-Boson Contribution}

When $f_2$ and $f_1$ are isospin partners, then there is a \textit{T-channel W-boson contribution}. Translating from \cite{3Christensen2024},
\begin{equation}\label{EQCA: W T-channel}
    \begin{aligned}
        \mathcal{M}_{WT}&= \kappa_{WT}2m_W^2\left[\hat{B}\widetilde{\lambda}_1\lambda_4\right]^{\jj\mm}\left[\hat{B}^\dagger\lambda_2^\dagger\lambda_3^-\right]^{\kk\Le} \\
        &+\kappa_{WT}\left(m_2^2\left[\hat{B}\widetilde{\lambda}_1\lambda_3\right]^{\jj\Le}-m_1m_2\left[\hat{B}^\dagger\lambda_1^\dagger\lambda_3^-\right]^{\jj\Le}\right)\left[\hat{B}^\dagger\lambda_2^\dagger\lambda_4^-\right]^{\kk\mm} \\
        &-\kappa_{WT}\left(m_1m_2\left[\hat{B}\widetilde{\lambda}_1\lambda_3\right]^{\jj\Le}-m_1^2\left[\hat{B}^\dagger\lambda_1^\dagger\lambda_3^-\right]^{\jj\Le}\right)\left[\hat{B}\widetilde{\lambda}_2\lambda_4\right]^{\kk\mm},
    \end{aligned}
\end{equation}
where $\kappa_{WT}=e^2/2m_W^2s_W^2(t-m_W^2)$. 

\subsection{$f\overline{f}hh$ Scattering Amplitudes}

For two massive fermions and two Higgs-bosons, there are contributions in the S-channel, T-channel, and U-channel for a total of three terms. The full amplitude is
\begin{equation}\label{EQCA: Full ffhh}
    \mathcal{M}_{f\overline{f}hh}=\mathcal{M}_{hS}+\mathcal{M}_{fT}+\mathcal{M}_{fU},
\end{equation}
and when squared there are squared terms and \textit{crossterms}. Both squared terms and crossterms have been confirmed to agree with the values calculated in both the CSM and the FSM.

For the upcoming constructive amplitude contributions, the Lorentz spinors $\lambda_1$ and $\lambda_2$ correspond to $f$ and $\overline{f}$, and the spacetime momenta $p_3$ and $p_4$ correspond to the Higgs-bosons.

\subsubsection{S-Channel Higgs Contribution}

Translating from \cite{3Christensen2024}, the \textit{S-channel Higgs contribution} is 
\begin{equation}\label{EQCA: S Higgs ffhh}
    \mathcal{M}_{hS}=\tau_{hS}\left(\left[\hat{B}\widetilde{\lambda}_1\lambda_2\right]^{\jj\kk}+\left[\hat{B}^\dagger\lambda_1^\dagger\lambda_2^-\right]^{\jj\kk}\right),
\end{equation}
for $\tau_{hS}=-3e^2m_f^2m_h^2/4m_W^2s_W^2(s-m_h^2)$.

\subsubsection{T-Channel Fermion Contribution}

Translating from \cite{3Christensen2024}, the \textit{T-channel fermion contribution} is
\begin{equation}\label{EQCA: T Ferm. ffhh}
    \begin{aligned}
        \mathcal{M}_{fT} &= 2m_f\tau_{fT}\left(\left[\hat{B}\widetilde{\lambda}_1\lambda_2\right]^{\jj\kk}+\left[\hat{B}^\dagger\lambda_1^\dagger\lambda_2^-\right]^{\jj\kk}\right) \\
        &+ \tau_{fT}\left(\left[\hat{B}\lambda_1^\dagger p_3^-\lambda_2\right]^{\jj\kk}+\left[\hat{B}^\dagger\widetilde{\lambda}_1p_3\lambda_2^-\right]^{\jj\kk}\right),
    \end{aligned}
\end{equation}
for $\tau_{fT}=\tau_{hS}(s-m_h^2)/3m_h^2(t-m_f^2)$.

\subsubsection{U-Channel Fermion Contribution}

Translating from \cite{3Christensen2024}, the \textit{U-channel fermion contribution} is
\begin{equation}\label{EQCA: U Ferm. ffhh}
    \begin{aligned}
        \mathcal{M}_{fU} &= 2m_f\tau_{fU}\left(\left[\hat{B}\widetilde{\lambda}_1\lambda_2\right]^{\jj\kk}+\left[\hat{B}^\dagger\lambda_1^\dagger\lambda_2^-\right]^{\jj\kk}\right) \\
        &+ \tau_{fU}\left(\left[\hat{B}\lambda_1^\dagger p_4^-\lambda_2\right]^{\jj\kk}+\left[\hat{B}^\dagger\widetilde{\lambda}_1p_4\lambda_2^-\right]^{\jj\kk}\right),
    \end{aligned}
\end{equation}
for $\tau_{fU}=\tau_{fT}(t-m_h^2)/(u-m_h^2)$, with $u$ being a U-channel Mandelstahm variable.

\printbibliography


\end{document}